\documentclass[twocolumn]{aastex63}

\usepackage{graphicx}	
\usepackage{amsmath}	
\usepackage{amssymb}	
\usepackage{xspace}
\usepackage{longtable}
\usepackage{scalerel}
\newcommand{\HI}{H\protect\scaleto{$I$}{1.2ex}\xspace}
\newcommand{\Ha}{H$\alpha$\xspace}

\newcommand{\hdue}{$\rm H_{2}$\xspace}
\newcommand{\Msun}{$\rm M_\odot$\xspace}

\newcommand{\kms}{$\rm km \, s^{-1}$\xspace}

\newcommand{\aco}{$\alpha_{CO}$\xspace}
\received{June 30, 2023}
\revised{August 8, 2023}
\accepted{August 6, 2023}

\shorttitle{\hdue and \HI in RPS galaxies}
\shortauthors{Moretti et al.}
\graphicspath{{./}{figures/}}

\begin{document}

\title{The evolution of the cold gas fraction in nearby clusters ram-pressure stripped galaxies}

\correspondingauthor{Alessia Moretti}
\email{alessia.moretti@inaf.it}

\author[0000-0002-1688-482X]{Alessia Moretti}
\affiliation{INAF-Padova Astronomical Observatory, Vicolo dell'Osservatorio 5, I-35122 Padova, Italy}

\author[0000-0001-5965-252X]{Paolo Serra}
\affiliation{INAF-Cagliari Astronomical Observatory, Via della Scienza 5, I-09047 Selargius (CA), Italy}

\author[0000-0002-8372-3428]{Cecilia Bacchini}
\affiliation{INAF-Padova Astronomical Observatory, Vicolo dell'Osservatorio 5, I-35122 Padova, Italy}

\author[0000-0001-9143-6026]{Rosita Paladino}
\affiliation{INAF-Istituto di Radioastronomia, via P. Gobetti 101, I-40129 Bologna, Italy}

\author[0000-0003-0231-3249]{Mpati Ramatsoku}
\affiliation{Department of Physics and Electronics, Rhodes University, PO Box 94, Makhanda, 6140, South Africa}
\affiliation{South African Radio Astronomy Observatory, 2 Fir Street, Black River Park, Observatory, Cape Town, 7405, South Africa}
\affiliation{INAF-Cagliari Astronomical Observatory, Via della Scienza 5, I-09047 Selargius (CA), Italy}

\author[0000-0001-8751-8360]{Bianca M. Poggianti}
\affiliation{INAF-Padova Astronomical Observatory, Vicolo dell'Osservatorio 5, I-35122 Padova, Italy}

\author[0000-0003-0980-1499]{Benedetta Vulcani}
\affiliation{INAF-Padova Astronomical Observatory, Vicolo dell'Osservatorio 5, I-35122 Padova, Italy}

\author[0000-0003-1078-2539]{Tirna Deb}
\affiliation{University of Western Cape, South Africa}

\author[0000-0002-7296-9780]{Marco Gullieuszik}
\affiliation{INAF-Padova Astronomical Observatory, Vicolo dell'Osservatorio 5, I-35122 Padova, Italy}

\author[0000-0002-7042-1965]{Jacopo Fritz}
\affiliation{Instituto de Radioastronomia y Astrofisica, UNAM, Campus Morelia, AP 3-72, CP 58089, Mexico}

\author[0000-0001-5840-9835]{Anna Wolter}
\affiliation{INAF-Osservatorio Astronomico di Brera, Via Brera, 28, I-20121 Milano, Italy}

\begin{abstract}
Cluster galaxies are affected by the surrounding environment, which influences, in particular, their gas, stellar content and morphology. In particular, the ram-pressure exerted by the intracluster medium promotes the formation of multi-phase tails of stripped gas detectable both at optical wavelengths and in the sub-mm and radio regimes, tracing the cold molecular and atomic gas components, respectively. In this work we analyze a sample of sixteen galaxies belonging to clusters at redshift $\sim 0.05$ showing evidence of an asymmetric \HI morphology (based on MeerKAT observations) with and without a star forming tail. To this sample we add three galaxies with evidence of a star forming tail and no \HI detection. Here we present the galaxies \hdue gas content from APEX observations of the CO(2-1) emission. We find that in most galaxies with a star forming tail the \hdue global content is enhanced with respect to undisturbed field galaxies with similar stellar masses, suggesting an evolutionary path driven by the ram-pressure stripping. As galaxies enter into the clusters their \HI is displaced but also partially converted into \hdue, so that they are \hdue enriched when they pass close to the pericenter, i. e. when they develop the star forming tails that are visible in UV/B broad bands and in \Ha emission. An inspection of the phase-space diagram for our sample suggests an anticorrelation between the \HI and \hdue gas phases as galaxies fall into the cluster potential. This peculiar behaviour is a key signature of the ram-pressure stripping in action.
\end{abstract}


\section{Introduction} \label{sec:intro}

Cluster galaxies are well known to display global properties which are different from those exhibited by galaxies located in the field: their integrated colors are redder, at fixed morphology; the distribution of morphological types is skewed toward a preponderance of early-type galaxies, at least where the local density is higher \citep{dressler80,Fasano2015,Vulcani+2023}; their Star Formation History often shows signatures of  quenching, both on long (the so-called starvation) and short timescales \citep{Guglielmo+2015,Paccagnella2016,Paccagnella+2017,Paccagnella+2019}.
All these properties are obviously correlated, but which transformation happens first and when is still a matter of debate.
Various physical mechanism can shape galaxy properties in dense environments \citep{BG06,Cortese+2021}, among which we recall the galaxy interaction through multiple encounters and interaction with the cluster potential \citep{Byrd+Valtonen1990,moore+1999},  
the thermal evaporation of the cold interstellar medium (ISM) at the interface with the hot intracluster medium \citep{CowieSongaila1977}, the exhaustion of the atomic gas fuel (starvation) caused  by the removal of the circum-galactic hot corona due to the interaction with the intracluster medium \citep{Larson+1980}, the removal of the gas from the star forming disk of late-type galaxies due to the viscosity momentum transfer with the intracluster medium \citep{Nulsen1982} or due to their infall toward the central region of the cluster. This last mechanism is called ram-pressure stripping (hereafter RPS) \citep{GG72}, and it has been extensively studied first through the radio continuum and 21-cm emission of the cold atomic gas \citep{Gavazzi1978}.
More recently, thanks to the availability of Integral Field Units operating at optical wavelengths,  the ionized gas emission (sometimes associated with the ongoing star formation) in the stripped gaseous tails originated from this mechanism 
has started to be studied in local and 
intermediate redshifts environments
\citep{Merluzzi2013,Fumagalli2014,Fossati2016,GASPI,Moretti+2022} .

In particular, the GASP\footnote{https://web.oapd.inaf.it/gasp/} survey \citep{GASPI} used 
MUSE integral field data of a large sample of galaxies in low redshift clusters showing evidences of RPS, allowing a statistically significant view on this mechanism.
The GASP sample of cluster galaxies has been selected on the basis of the optical morphology, i. e. mostly relying on the asymmetries evidenced by the B-band emission in WINGS/OmegaWINGS images \citep{Poggianti+2016}, and is therefore biased towards galaxies with an optical star forming tail. In fact, most GASP galaxies have star forming tails, where the star formation takes place in knots, recently resolved by HST data \citep{Gullieuszik+2023,Giunchi+2023}.
MUSE data have also confirmed that, together with the appearance of the optical tail, galaxies experience an enhanced star formation in the disk \citep{Vulcani+2018}.
At the same time, GASP galaxies with an \HI mass determination \citep{GASPXVII,GASPXXVI,GASPXXXIX} have shown a short \HI depletion time in the disk. Unfortunately, the spatial resolution of our \HI VLA data does not allow a mapping of the depletion time, but only the derivation of a global value.
Before the advent of MUSE, in fact, RPS was already a well known mechanism, as \HI dedicated studies of nearby clusters already demonstrated how this phenomenon is able to produce galaxies with asymmetric disks, sometimes leading to cold gas tails and to almost completely stripped disks \citep{Kenney+1989,Kenney2004,Chung+2007,Cortese+2010,Hess+2022}.
Further evidence of stripping have been also detected  using X-ray emission \citep{Sun+2010,Jachym2017,Campitiello+2021,Bartolini+2022} and low frequency data \citep{Gavazzi1978,Gavazzi+1995,Roberts+2021,Ignesti+2022}.
In fact, the ability to detect RPS in action selecting galaxies on the basis of their optical appearance was somewhat unexpected until the advent of the MUSE IFU, able to detect all the optical emission lines needed to identify star forming regions in the gaseous tails.
In the last 20 years, therefore, the RPS studies shifted from searching for peculiar objects with asymmetric \HI morphologies in nearby clusters to characterizing their overall incidence, and to study the complex interplay between the various gas phases  \citep{GASPXXIII}.
The first of these studies were carried out in the nearby Virgo Cluster \citep{Corbelli+2012,Boselli+2014}, and it is recently the subject of a dedicated observational effort to detect both the ionized gas properties \citep{Boselli+2018} and the molecular gas distribution and properties with ALMA data \citep[VERTICO survey]{Brown+2021}, coupling them with resolved \HI from the VIVA survey \citep{Chung2009,Yoon+2017}. Resolved and deep observations on Virgo are being complemented nowadays with the ViCTORIA survey with MeerKAT \citep{Boselli+2023}.
Virgo galaxies have been confirmed to be subject to \HI and \hdue depletion, albeit with different efficiency: when looking at the spatially resolved properties, it turns out that while the \HI gets more easily displaced, the \hdue seems to be more centrally concentrated, suggesting that it is stripped as well in the disk outskirts \citep{Zabel+2022,Villanueva+2022}.
However, in Virgo galaxies the overall molecular gas content is not dramatically different from field galaxies, but galaxies showing evidence of ram pressure stripping seem to be   slightly enriched in \hdue \citep{Zabel+2022}.

Besides Virgo, the multiphase ISM is being studied in the other two nearest clusters Fornax \citep{Loni+2021,Kleiner+2021,morokuma-matsui+2022,Serra+2023,Kleiner+2023} and Coma \citep{RobertsParker2020,Molnar+2022}, using datasets with an homogeneous observational coverage over a wide field of view, finding results that bear a dependence on the hosting cluster mass. 
Fornax shows a slight \HI deficiency corresponding to a slight SFR decrease, whereas in Coma galaxies seem to be simultaneously \HI deficient and SFR enhanced.
In both cases, though, the molecular gas content of cluster galaxies is not yet firmly established.

However, APEX/ALMA data do exist for some of the GASP galaxies and they show the global \hdue content to be enhanced with respect to field galaxies having a comparable stellar mass \citep{GASPX,Moretti_2020}, leading to the suggestion that the part of the \HI gas is efficiently converted into \hdue.
In fact, an increase in the gas pressure, adding to the hydrostatic midplane one, is expected to be observed where the RP hits the galaxy disk, and may cause such an efficient conversion of the atomic cold gas in the molecular phase  \citep{Elmegreen1993,Wong+Blitz2002,Leroy2008,Krumholz+2009,Sun+2020}. Alternatively, the gas compression could follow the gas motion induced by the RP favoring the conversion along the direction of the stripping. Both these effects will be strongly dependent on the geometry of the stripping, and could only be analyzed using spatially resolved observations.

It is worth noticing, though, that GASP galaxies analyzed so far are all in the so called "peak stripping" phase, i.e. they have been caught at their pericentric passage, where the optical star forming tail is more evident. 
Their molecular gas content may be enhanced for a correspondingly short timescale (i.e. for the short period they are at the pericenter).
The question which naturally arises is whether the different gas phases trace an evolutionary sequence in the ram pressure mechanism, which implies that mapping together galaxies with a different degree of stripping in optical/\HI/\hdue, is possibly evidencing the typical signature of this effect.
We therefore analyze in this paper a sample of 19 galaxies showing evidence of stripping in \HI/optical, with the aim of deriving the temporal sequence of the ram pressure stripping.
We describe the target selection in Sec.\ref{sec:targets}, we then analyze APEX data in Sec.\ref{sec:data} and report on the cold gas content of these galaxies in Sec.\ref{sec:gascontent}. We then analyze the cold gas components (\HI and \hdue) in Sec. \ref{sec:coldgasratio}
and we draw our conclusions in Sec.\ref{sec:conclusions}.

 Throughout this paper we make use of the standard cosmology $H_0 = 70 \, \rm km \, s^{-1} \, Mpc^{-1}$, ${\Omega}_M=0.3$
and ${\Omega}_{\Lambda}=0.7$ and we express stellar masses using the Chabrier Initial Mass Function \citep[IMF,][]{Chabrier2003}.

\section{Targets selection}\label{sec:targets}

In order to select our targets we needed a good \HI and optical coverage with a wide field of view. We therefore matched the OmegaWINGS cluster catalog \citep{Gullieuszik+2015,Moretti+2017} with the MeerKAT Galaxy Cluster Legacy Survey \citep[MGCLS;][]{Knowles+2022} and we ended up with the three clusters A85, A3376 and A3558 for which we possess the \HI emission datacubes and the spectroscopic coverage. 
The analysis of the MeerKAT data will be described in Sec.~\ref{sec:HI}.
We selected all galaxies in these 3 clusters whose \HI contours were asymmetric, as, given the low physical resolution ($\sim$30 kpc), we consider these asymmetries as candidate tails of stripped gas. For simplicity, in the rest of this paper we refer to the asymmetries at all wavelengths, including those at 21 cm, as tails. A posteriori, we checked whether there was evidence for a coincident  stellar tail either in the B-band \citep[from][]{Gullieuszik+2015,Poggianti+2016} or in the UV \citep[from][ and George et al. in prep., using UVIT on AstroSat observations]{George+2023}, when available. 

We finally excluded from the sample those galaxies having a stellar mass below 10$^{10}$ \Msun, as they would require much deeper observations to detect a possible \hdue deficiency. 
We ended up with a sample of 16 targets. To this we added one galaxy showing a truncated disk morphology from the \Ha emission (A3376\_JW108) and two more galaxies belonging to clusters not yet covered by MeerKAT observations, which we selected because of their clear star-forming tails based on GASP MUSE data. These two galaxies are JO49 in A168 and JO60 in A1991 and have been both classified as jellyfishes by GASP, meaning that they have an ionized gas tail at least as long as the stellar disk. For them, our analysis is limited to the molecular gas content, but cannot include the \HI content.

Tab. \ref{tab:targets_data} contains the list of the observed targets as well as their optical coordinates, spectroscopic redshifts, and stellar masses.
The stellar masses have been derived from the existing optical spectroscopy either as fiber spectroscopy from the WINGS/OmegaWINGS spectra, corrected for aperture effects to a total mass,  or from the IFU integral within the galaxy disks for the galaxies that possess MUSE data from GASP \cite{Vulcani+2018}. When none of these information were available we used the stellar masses derived in \cite{Vulcani+2022_BG} using the optical photometry.
We also give in Col. 6 of Tab.\ref{tab:targets_data} the flag indicating whether the galaxy possess a star forming tail or not: we defined as star forming any tail with UV/\Ha or B-band emission (where the \Ha comes from the GASP data, when available), while we tagged as non star-forming those asymmetric \HI emission without any evidence of young stars in the tail direction.
Tab.\ref{tab:targets_data} also contains
the GASP name, when available, the classification of the galaxy according to the MUSE optical data (Poggianti+, in prep.) and the MeerKAT identification and \HI mass. 
Finally, the last column refers to the \HI morphology classified according to \citet{Yoon+2017}, that we will discuss in Sec.\ref{sec:coldgasratio} where we relate the properties of the two cold gas phases.

\begin{table*}
\movetabledown=3.5cm
\begin{rotatetable}
\caption{Targets id, coordinates of the pointings, redshift, stellar masses, tail star forming flag, corresponding GASP ID, GASP stripping stage and MeerKAT ID when available, \HI mass with errors in log units}. The last column shows the \HI classification according to \citep{Yoon+2017}.\label{tab:targets_data}
\begin{center}
\begin{tabular}{|l|c|c|c|c|c|l|l|c|c|}
\hline
Galaxy & RA & DEC & z  & log(M$_\star$) & SF  & GASP ID/Stripping stage& MKTCS-HI id&log(M$_{HI}$)  & \HI class\\
 &  &  &   &  & tail &  & & &\\
\hline
A168{\_}JO49	& 01:14:43.85 & +00:17:10.1 & 0.0453 & 10.7 & y & JO49, Peak (Jellyfish)&&&\\ 
A1991{\_}JO60	& 14:53:51.57 & +18:39:06.4 & 0.0621 & 10.4 & y & JO60, Peak (Jellyfish)&&&\\
A3376{\_}JW108	& 06:00:47.96 & -39:55:07.4 & 0.0479 & 10.5 & y & JW108, Advanced (Trunc. disk)  &&& None\\
A3376{\_}S45	& 06:02:41.30 & -40:14:57.2 & 0.0454 & 10.3 & y & &J060241.16-401458.5& 9.6 $^{+0.01}_{-0.05}$ & I \\
A3376{\_}S49   & 06:00:54.17 & -39:39:54.7 & 0.0465  & 10.3 & y & &J060053.89-393954.9& 9.2 $^{+0.04}_{-0.08}$ & II \\
A3376{\_}S55   & 06:02:55.44 & -40:22:11.0 & 0.0469  & 10.3 & y & &J060255.77-402211.0& 9.6 $^{+0.01}_{-0.06}$ & I\\
A3376{\_}S54   & 05:59:58.37 & -40:00:34.8 & 0.0469  & 10.0 & y & &J055958.48-400039.0& 8.8 $^{+0.11}_{-0.09}$ & IV\\
A3376{\_}S58   & 06:01:42.91 & -39:56:39.7 & 0.0473  & 10.5 & y & &J060142.83-395642.5& 9.0 $^{+0.07}_{-0.08}$ &IV\\
A3376{\_}S64   & 06:01:06.96 & -39:55:01.8 & 0.0480  & 10.1 & y & &J060107.48-395458.7& 9.4 $^{+0.02}_{-0.05}$ &II\\
A3376{\_}S74   & 06:02:07.09 & -39:42:30.4 & 0.0504  & 10.7 & n & &J060207.27-394226.5& 9.2 $^{+0.03}_{-0.06}$ &II\\
A3376{\_}S75   & 06:00:13.66 & -39:34:48.4 & 0.0509  & 11.0 & n& A3376\_B\_0261, Control & J060013.80-393449.6& 9.6 $^{+0.01}_{-0.05}$ &I\\
A3376{\_}S84   & 06:00:47.32 & -40:19:45.0 & 0.0533  & 11.0 & n & &J060047.86-401953.1& 9.2 $^{+0.05}_{-0.09}$ &III\\
A3558{\_}S167  & 13:26:49.75 & -31:23:44.5 & 0.0506  & 11.0 & y & JO147, Peak (Jellyfish)& J132649.98-312336.6& 9.1 $^{+0.05}_{-0.08}$ &II\\
A3558{\_}S60   & 13:28:18.24 & -31:48:17.1 & 0.0449  & 10.1 & y & JO157, Initial & J132817.83-314814.0& 9.5 $^{+0.02}_{-0.06}$ &I\\
A3558{\_}S124   & 13:26:35.67 & -30:59:36.3 & 0.0482  & 10.2 & y & JO159, Initial & J132635.48-305935.6& 9.5 $^{+0.02}_{-0.07}$ &I\\
A3558{\_}S134   & 13:29:28.54 & -31:39:25.4 & 0.0481  & 10.1 & y & JO160, Peak (Jellyfish) & J132927.44-313922.8& 9.3 $^{+0.03}_{-0.07}$ &II\\
A85{\_}S57     & 00:40:31.65 & -09:13:19.8 & 0.0512  & 11.2 & n & &J004031.89-091318.2& 9.5 $^{+0.02}_{-0.07}$ &II\\
A85{\_}S64     & 00:42:05.03 & -09:32:04.0 & 0.0529  & 10.9 & n & JO200, Unwinding/face-on& J004205.02-093206.5& 9.8 $^{+0.01}_{-0.05}$ &I\\
A85{\_}S90     & 00:42:31.94 & -09:17:50.4 & 0.0076* & 10.4 & n && J004231.46-091746.1& 9.2 $^{+0.05}_{-0.09}$ &II\\ 
\hline
\end{tabular}
\end{center}
\end{rotatetable}
\end{table*}

\begin{figure*}
    \centering
    \includegraphics[width=0.3\textwidth]{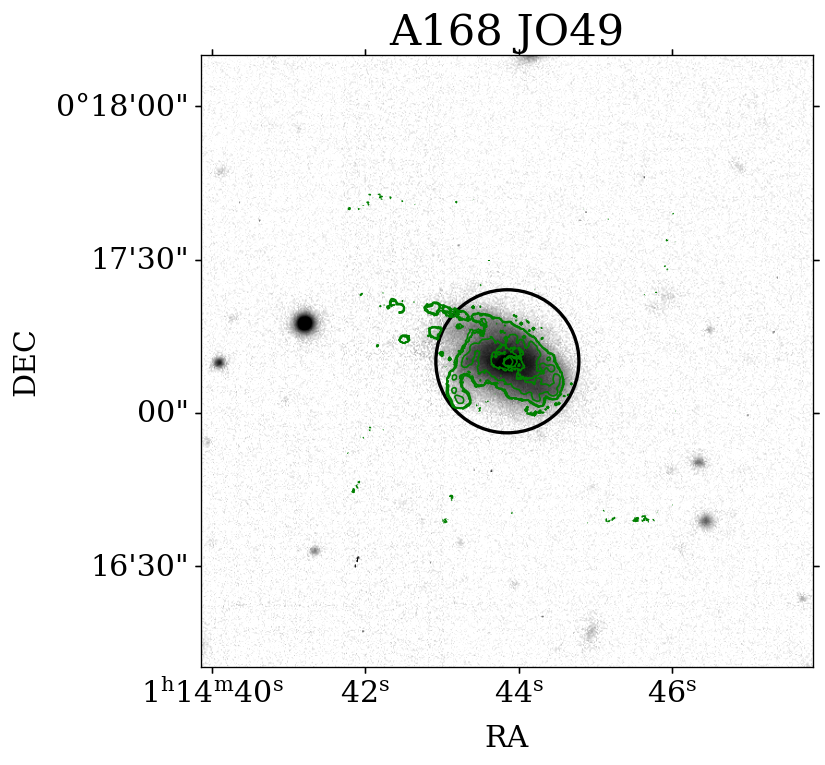} 
    \includegraphics[width=0.3\textwidth]{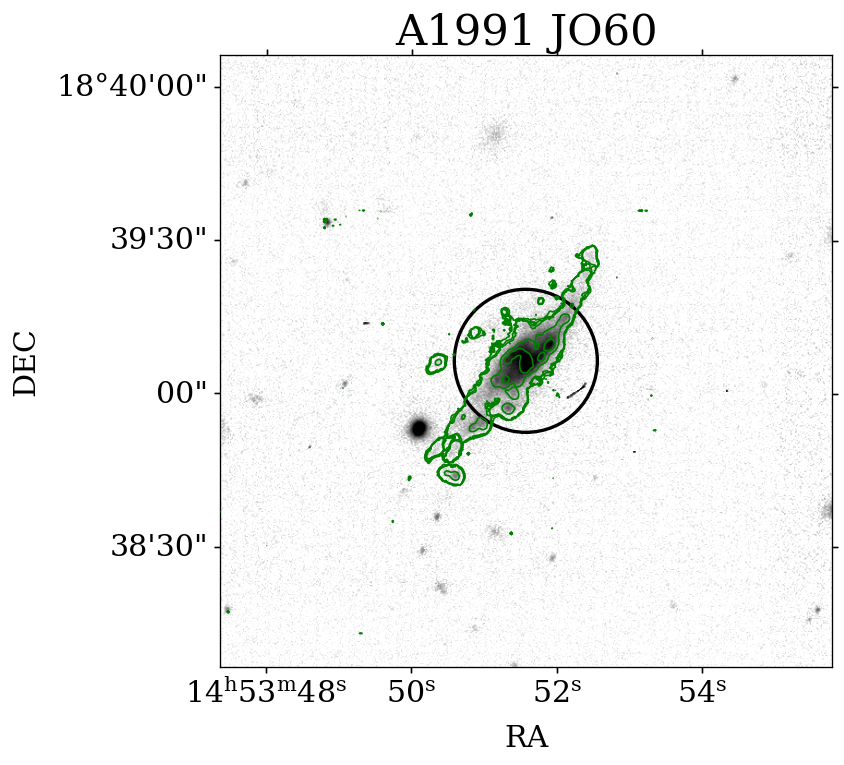}
    \includegraphics[width=0.3\textwidth]{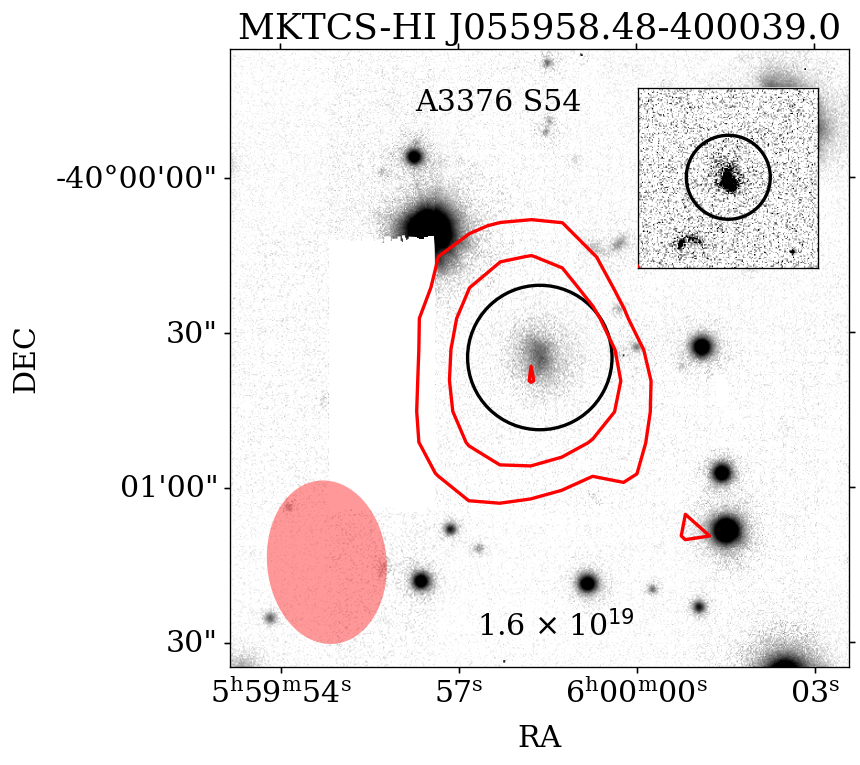}
    \includegraphics[width=0.3\textwidth]{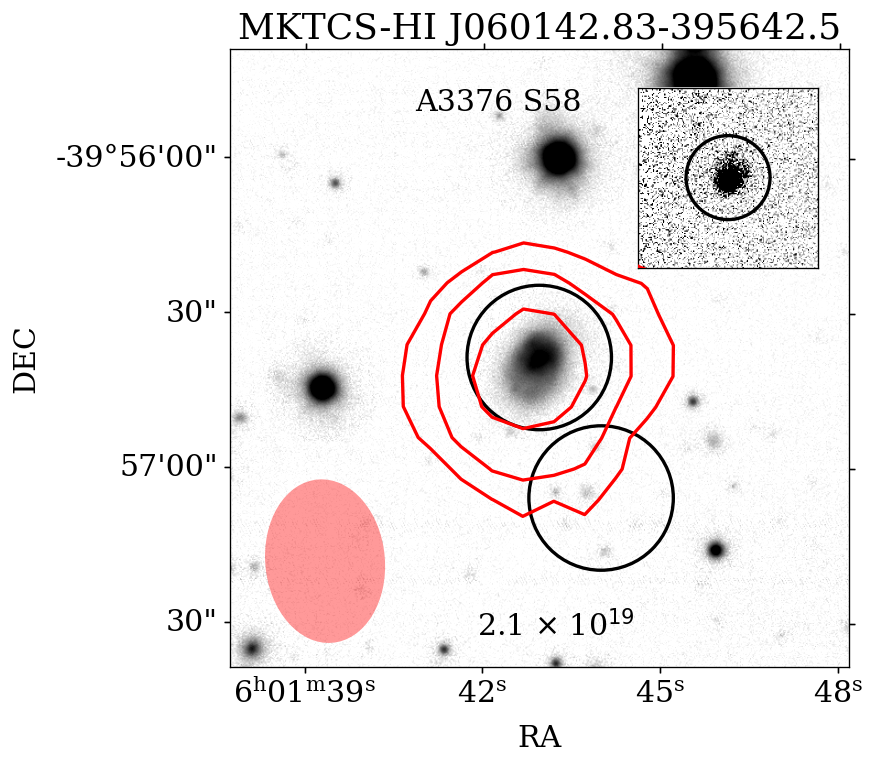}
    \includegraphics[width=0.3\textwidth]{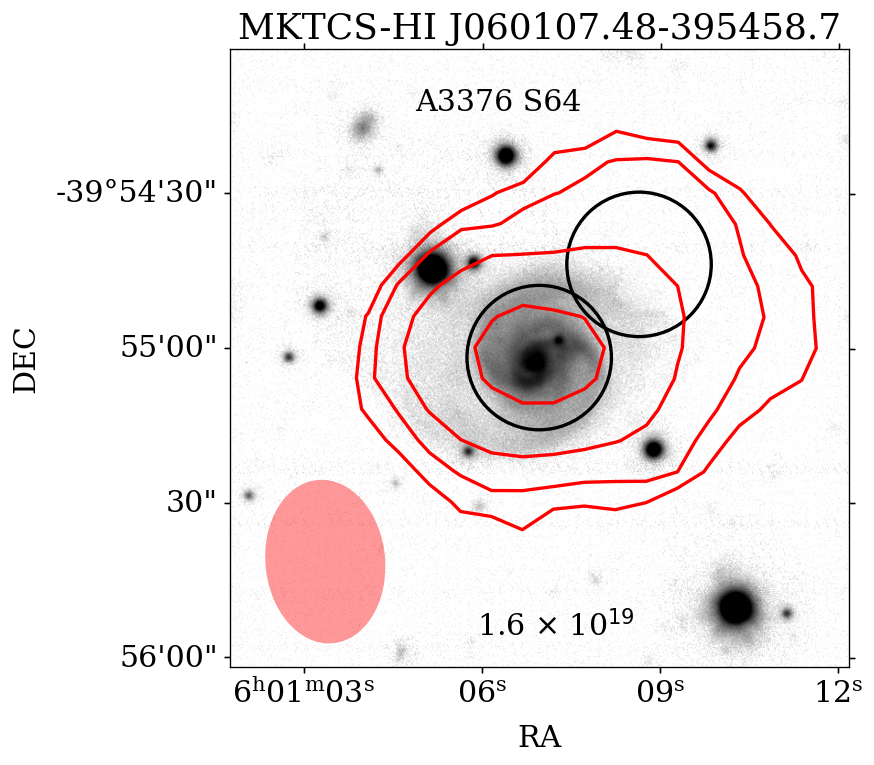}
    \includegraphics[width=0.3\textwidth]{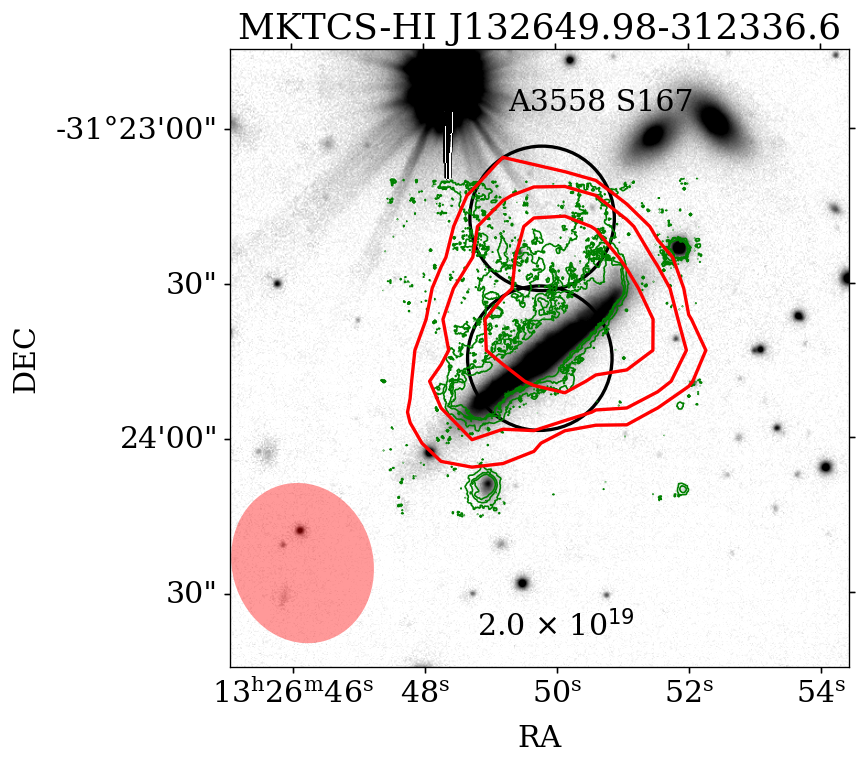}
    \includegraphics[width=0.3\textwidth]{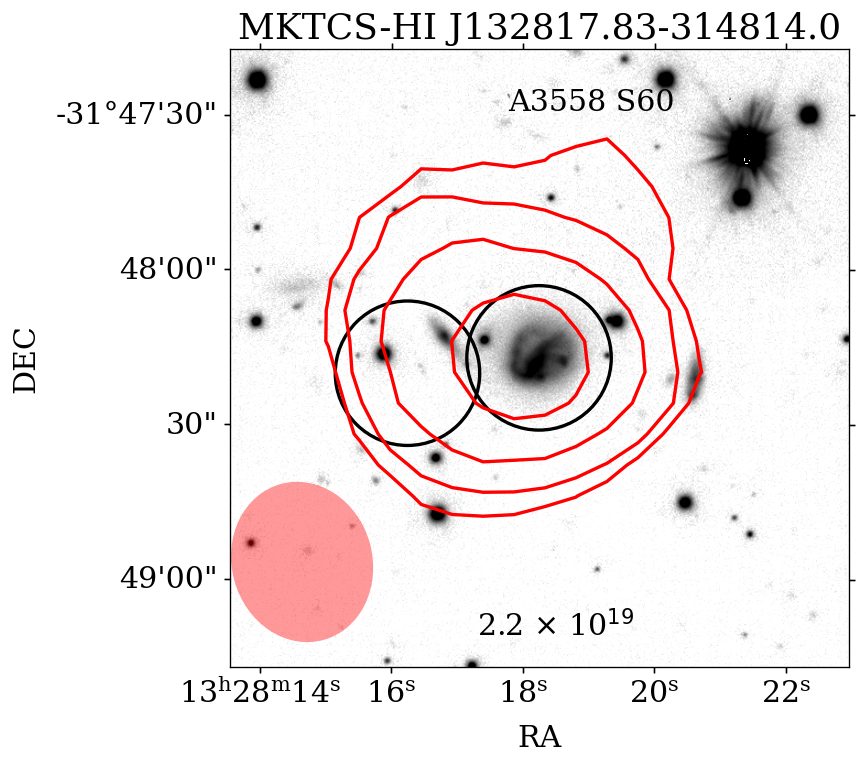}
    \includegraphics[width=0.3\textwidth]{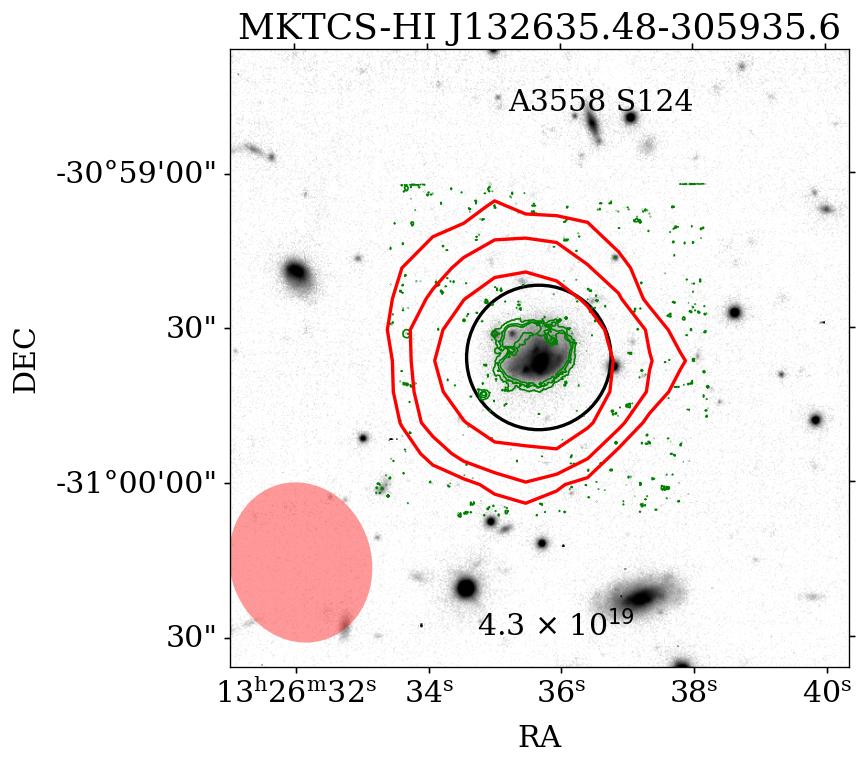}
    \includegraphics[width=0.3\textwidth]{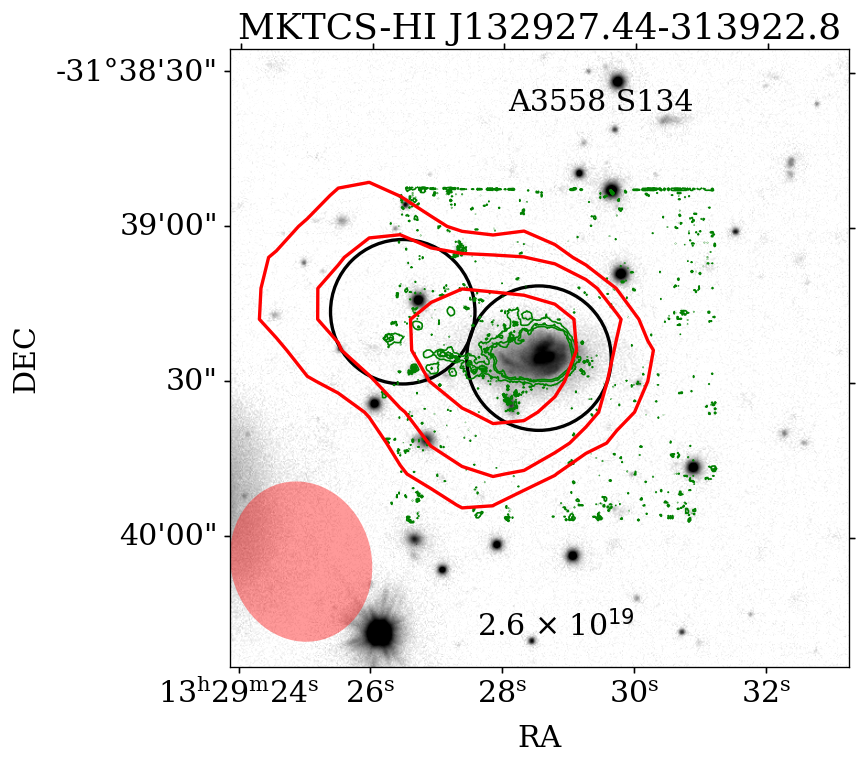}
    \includegraphics[width=0.3\textwidth]{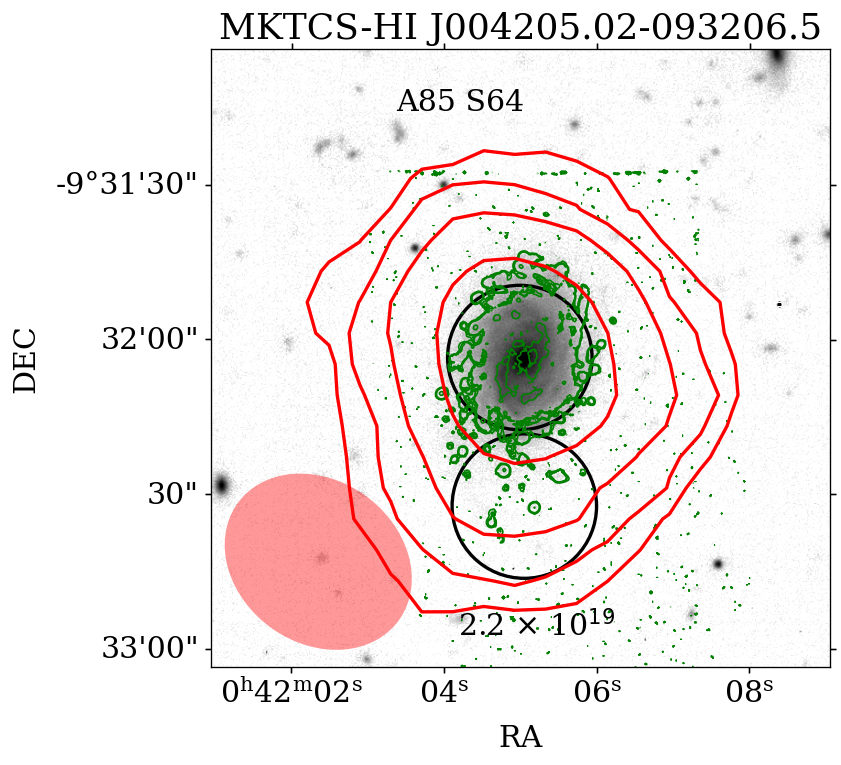}
     \caption{Optical maps from OmegaWINGS \citep{Gullieuszik+2015} for the galaxies that have been classified as having a star forming tail. The red contours show the radio \HI emission from MeerKAT corresponding to 1,2,4,8 $\times$ the lowest column density level (in atoms cm$^{-2}$) shown within each map. The red ellipse in the lower left corner is the MeerKAT beam. The greyscale insets show the UVIT data, when available. Black circles in each map and inset are the APEX pointings. Green contours show the ionized gas emission from MUSE, when available. }
    \label{fig:maps_sf}
\end{figure*}

\begin{figure*}
    \centering
    \includegraphics[width=0.3\textwidth]{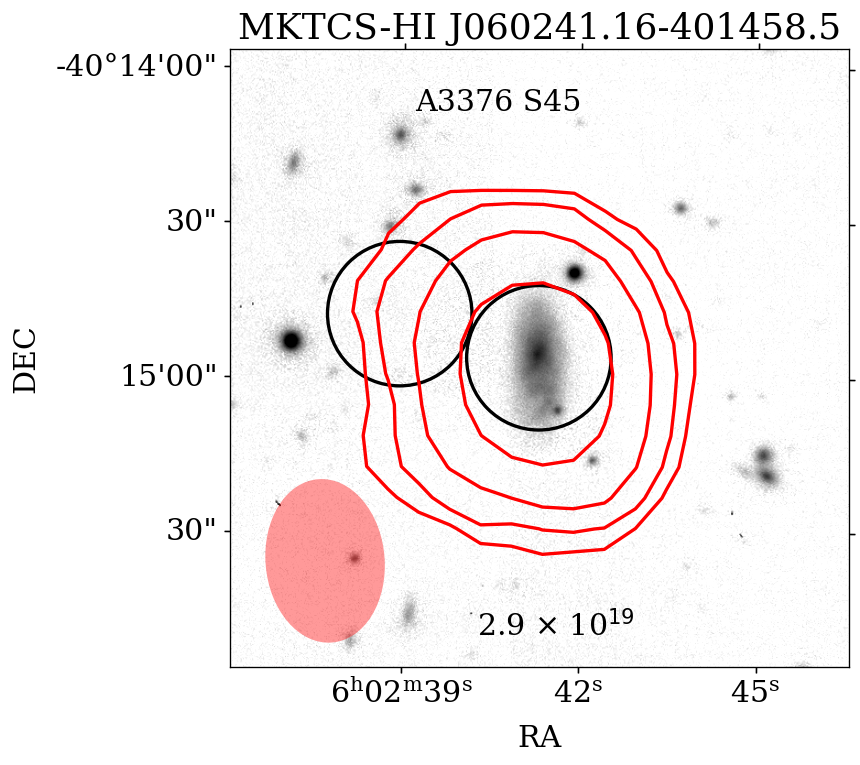}
    \includegraphics[width=0.3\textwidth]{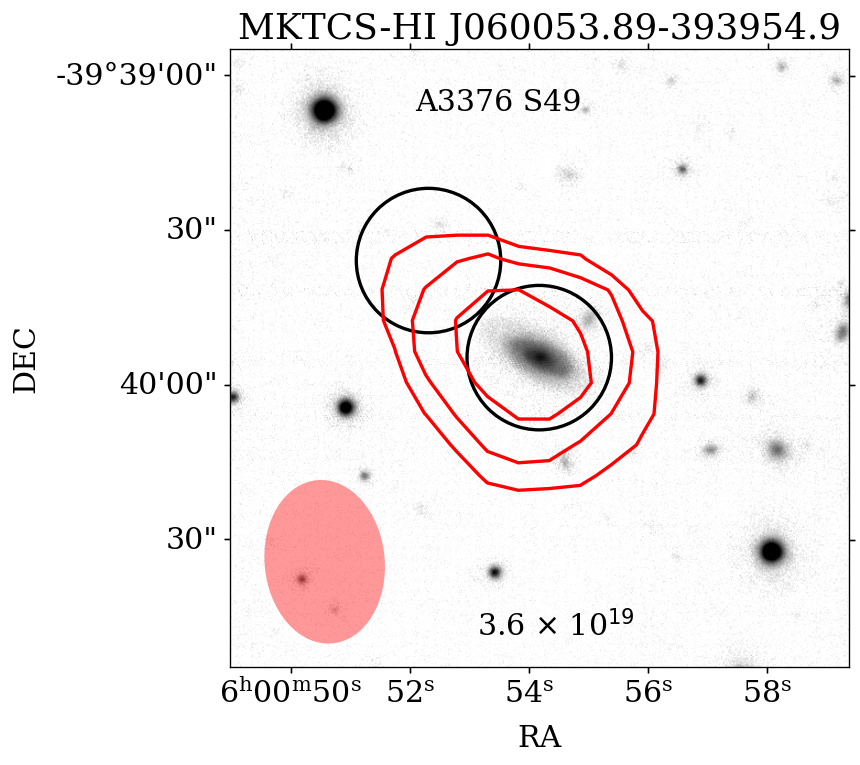}
    \includegraphics[width=0.3\textwidth]{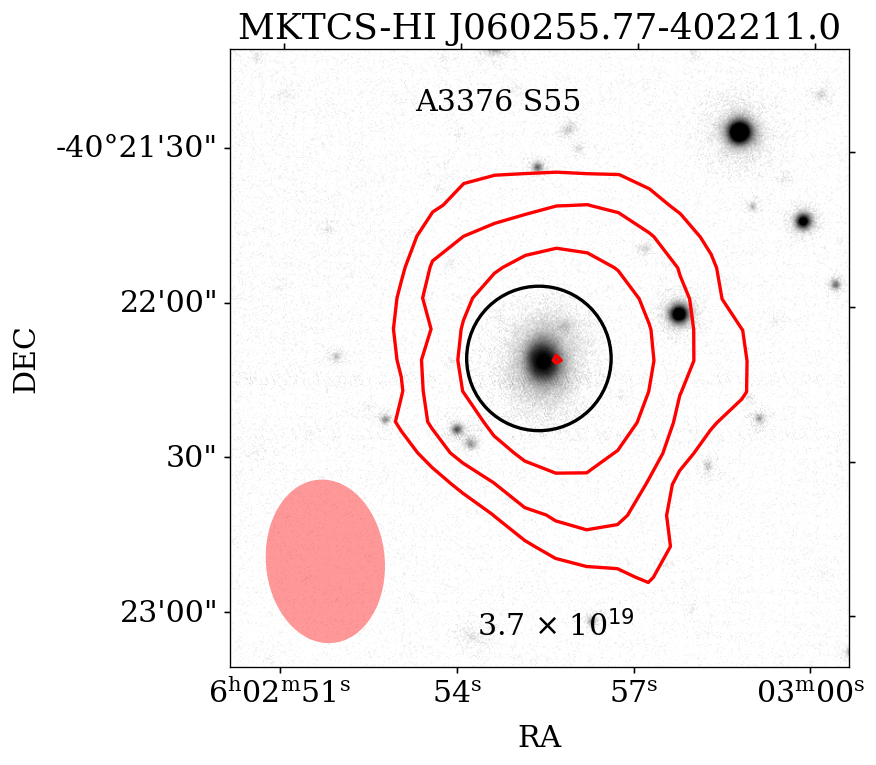}
    \includegraphics[width=0.3\textwidth]{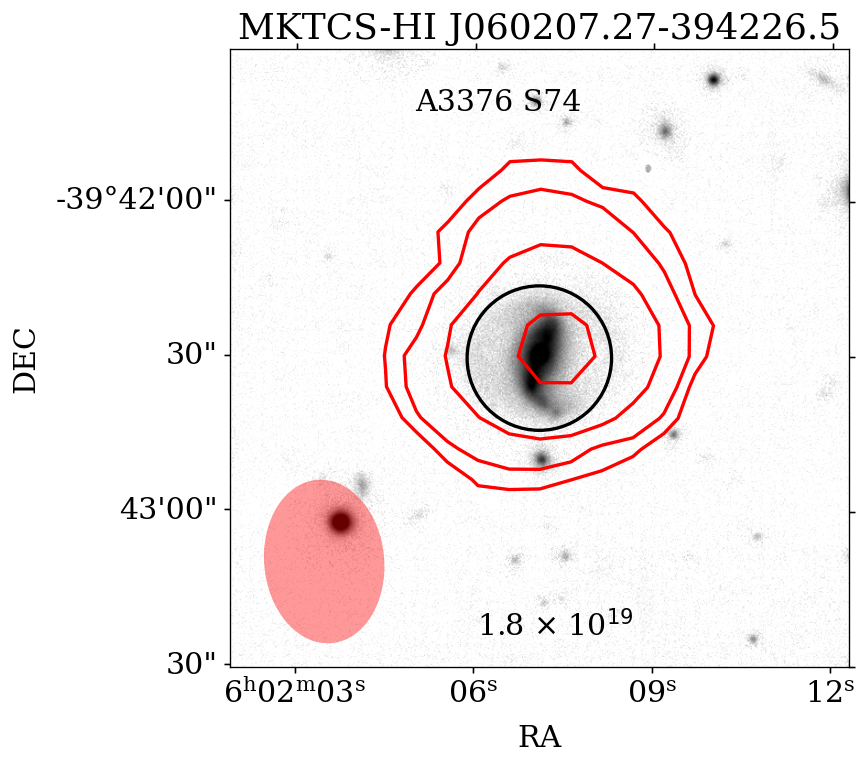} 
    \includegraphics[width=0.3\textwidth]{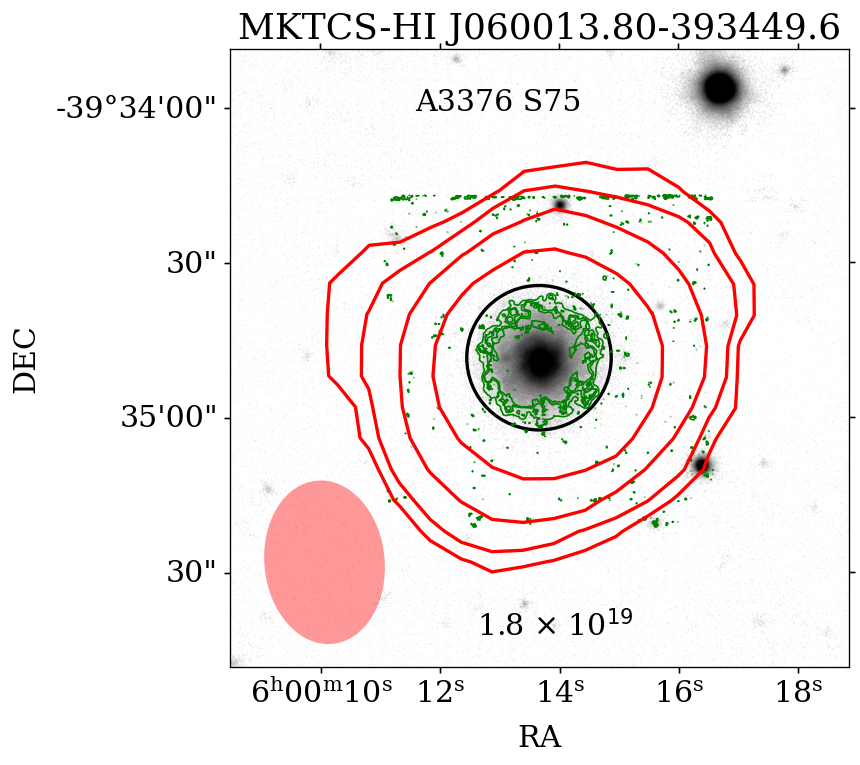}
    \includegraphics[width=0.3\textwidth]{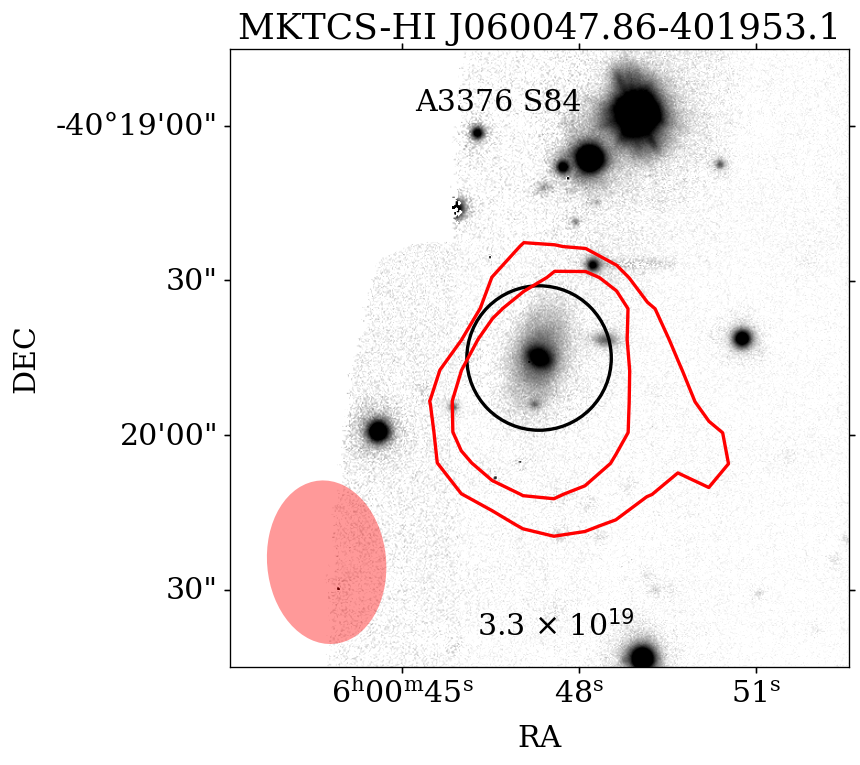}
    \includegraphics[width=0.3\textwidth]{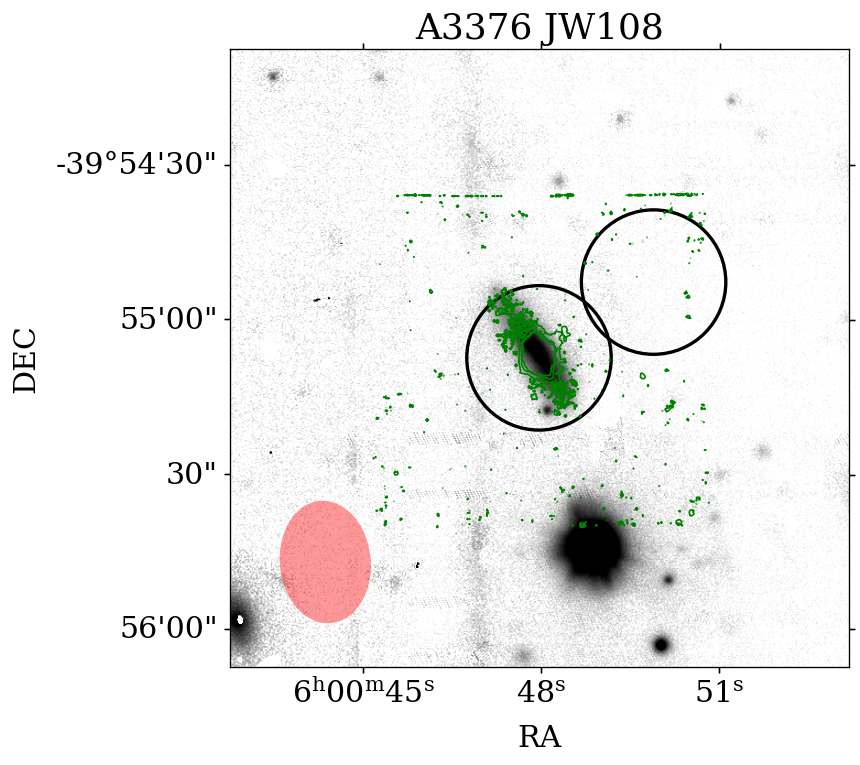}
    \includegraphics[width=0.3\textwidth]
    {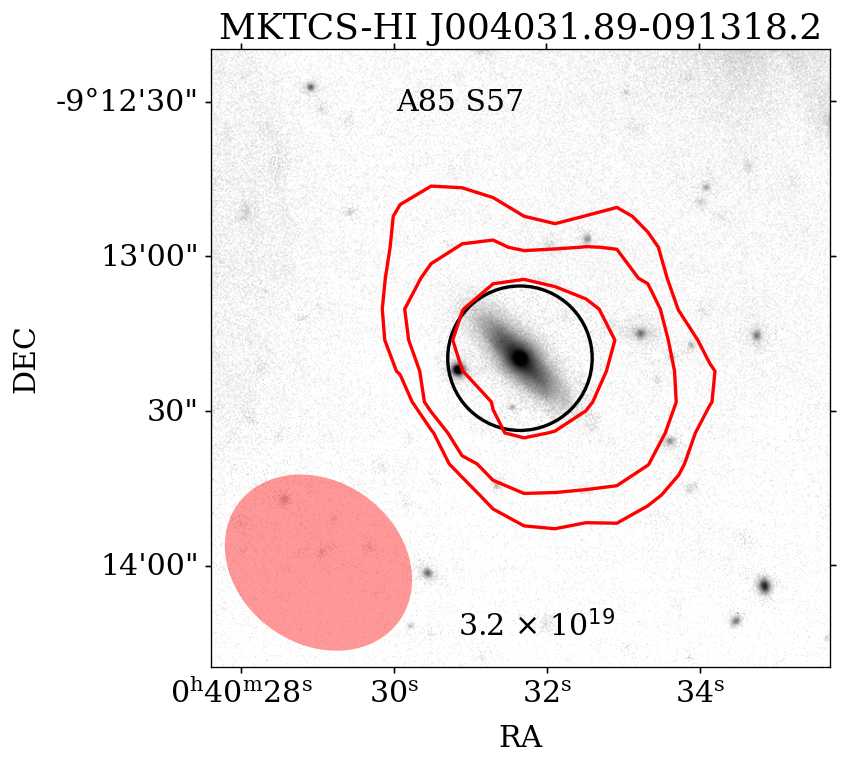}
    \includegraphics[width=0.3\textwidth]{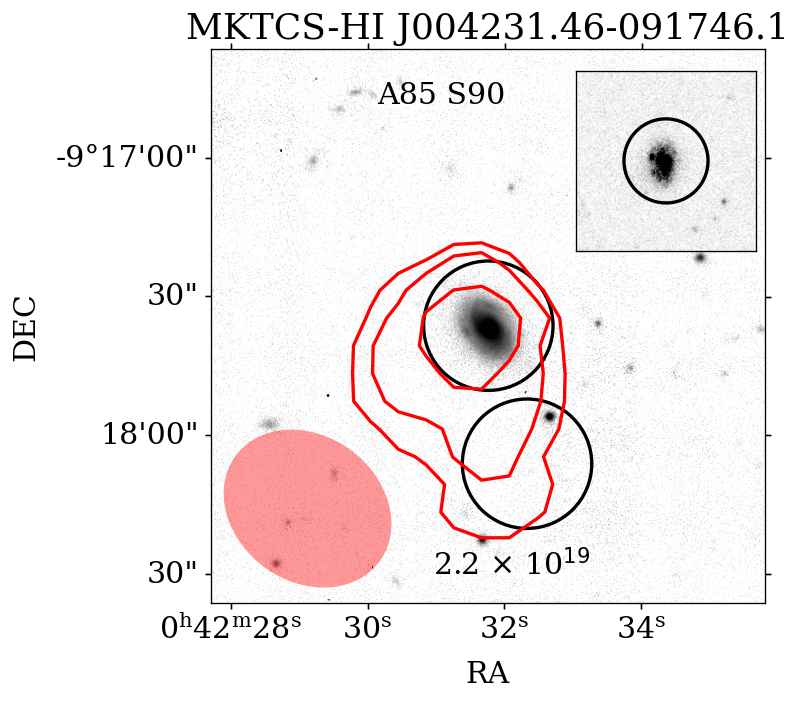}
     \caption{Optical maps from OmegaWINGS \citep{Gullieuszik+2015} for the galaxies that have been classified as having no star formation in the tail. Colors and symbols as in Fig.~\ref{fig:maps_sf}.}
    \label{fig:maps_nsf}
\end{figure*}

The optical maps (2\arcmin x 2\arcmin) of all our targets are shown in Fig.~\ref{fig:maps_sf} and Fig.~\ref{fig:maps_nsf}, representing galaxies with and without a star forming tail, respectively.
Black circles in Fig.~\ref{fig:maps_sf} and  Fig.~\ref{fig:maps_nsf} show the APEX pointings. At the observed frequencies (220 GHz) the FWHM of the APEX  primary beam is 28 \arcsec, which corresponds to scales from $\sim$25 to $\sim$30 kpc for the different targets. For 8 targets out of the 19 observed we also have a dedicated APEX pointing centred on the supposed emitting gas tail, which is also shown in Fig.~\ref{fig:maps_sf} and Fig.~\ref{fig:maps_nsf}.

\section{Data}\label{sec:data}

\subsection{HI data}\label{sec:HI}
We used the MeerKAT observations from the MeerKAT Galaxy Cluster Legacy Survey \citep[MGCLS;][]{Knowles+2022}. The MGCLS consists of long-track (6–10 hour) observations of 115 galaxy clusters in the Southern Sky using the L-band receiver (900 – 1670 MHz). The observed band was sampled with the 4k mode of the SKARAB correlator with 4096 channels that are 209 kHz ($\sim$44 \kms) wide in total polarisation. 

We processed the data over the frequency range, covering clusters A85, A3376 and A3558, and their background, $\sim$ 1300 MHz - 1410 MHz, centred at 1356 MHz. The uv-data were reduced using the standard procedure with the Containerised  Automated Radio Astronomical Calibration \citep[CARAcal;][]{Jozsa+2020} pipeline\footnote{https://caracal.readthedocs.io}. It is built using a radio interferometry framework, \textsc{stimela}\footnote{https://github.com/SpheMakh/Stimela} \citep{makhathini2018}, which is based on container technologies and Python. Various open-source radio interferometry software packages within this framework are available in the same script, such as CASA \citep{CASA}, WSclean \citep{wsclean,wsclean2}, AOFlagger \citep{aoflagger}, Cubical \citep{cubical}, SoFiA \citep{sofia,sofia2}, PyBDSF \citep{pybdsf}, MeqTrees \citep{meqtrees} and Crystalball\footnote{https://github.com/caracal-pipeline/crystalball}.

The uv-data were Fourier transformed into a final \HI cube with a pixel size of 6\arcsec\ and a field of view (FOV) of 2 deg$^{2}$. We used natural weighting with \textit{Briggs} robust parameter, r = 0.5 and a uv-tapering of 20\arcsec\ to optimise the surface brightness sensitivity. The resulting cubes have a rms of $\sigma =$ 0.2 mJy/beam. 
The sidelobes of the synthesised beam were removed by iteratively using \textit{SoFiA} to produce 3D clean masks and imaging with \textit{WSclean} while cleaning within the masks down to 0.5$\sigma$. The restoring Gaussian PSF of each cube has an average FWHM of ($\theta_{\rm maj} \times \theta_{\rm min}$) $\approx$  30\arcsec\ $\times$ 33\arcsec\ with a position angle, PA $\sim$150 deg. With these data we reach a column density sensitivity of n$_{\rm \HI} = 2.9 \times 10^{19}$ atoms cm$^{-2}$ at 3$\sigma$ assuming a linewidth of 44 \kms.

We then explored the final \HI cube using the \textit{SoFiA} source finder \citep{sofia} with the smooth and clip (S+C) \textit{SoFiA} method, searching for \HI emission brighter than 3$\sigma$.

The \HI masses of detected galaxies were calculated using the following:   

\begin{flushleft}
\begin{equation}\label{HImassEq}
\mathrm{M_{\rm \HI} = 2.36 \times 10^{5}D^{2}\int S_{\rm v} dv},
\end{equation}
\end{flushleft}

where the total integrated flux,  Svdv is expressed in Jy \kms, and D is the distance to the galaxy in Mpc. We adopt the luminosity distance using the optical velocities of the clusters, i.e. $D_L=245, 212, 202$ Mpc for A85, A3558 and A3376, respectively.
The derived \HI masses are shown in the 8$^{th}$ column of Tab.\ref{tab:targets_data}.

\subsection{APEX data}
The APEX data have been collected during ESO P108 under the program 0108.A-0511A (P. I. Moretti), that was awarded 184 hours of observational time in priority B.
Observations have been carried out during the period from August 2021 to December 2021, and the total observational time spent on the requested targets was 88.7 hours.
Each source has been observed to the requested noise level limit which is dependent on the galaxy stellar mass, and goes from 0.14 mK for the low mass galaxies to 0.25 mK for the high mass ones. These limits correspond to molecular gas fractions 0.5 dex lower than typical field galaxies \citep{Saintonge2017}.
We observed the $^{12}\rm CO(2-1)$ transition ($\nu_{rest}=230.538$ GHz), using the nFLASH230  Instrument tuned to the CO line redshifted frequencies for each target. 
The observations have been performed in a symmetric Wobbler switching mode, with maximum separation between the ON and OFF-beam of 100 arcsec. 

The spectra calibrated by the APEX on-line calibration pipeline (in antenna temperature scale; T$_{\rm A}^{*}$), have been checked and analysed using different spectral resolutions, going from 10 to 150 km/s.
First-order baselines, defined in a line-free band about 2000 km/s wide, have been subtracted 
from the spectra.
The antenna temperatures have been converted to main-beam brightness temperatures 
(T$_{\rm mb}$=T$_{\rm A}^{*}$/$\eta_{\rm mb}$), using $\eta_{\rm mb}$=0.75.

As shown in Fig.~\ref{fig:maps_sf} and Fig.~\ref{fig:maps_nsf}, the APEX pointing is encompassing the entire galaxy stellar extent in most cases, and we therefore are confident that with our measurements we are not missing any significant cold gas detection within the stellar disk. 

\begin{figure*}
    \centering
    \includegraphics[width=0.22\textwidth]{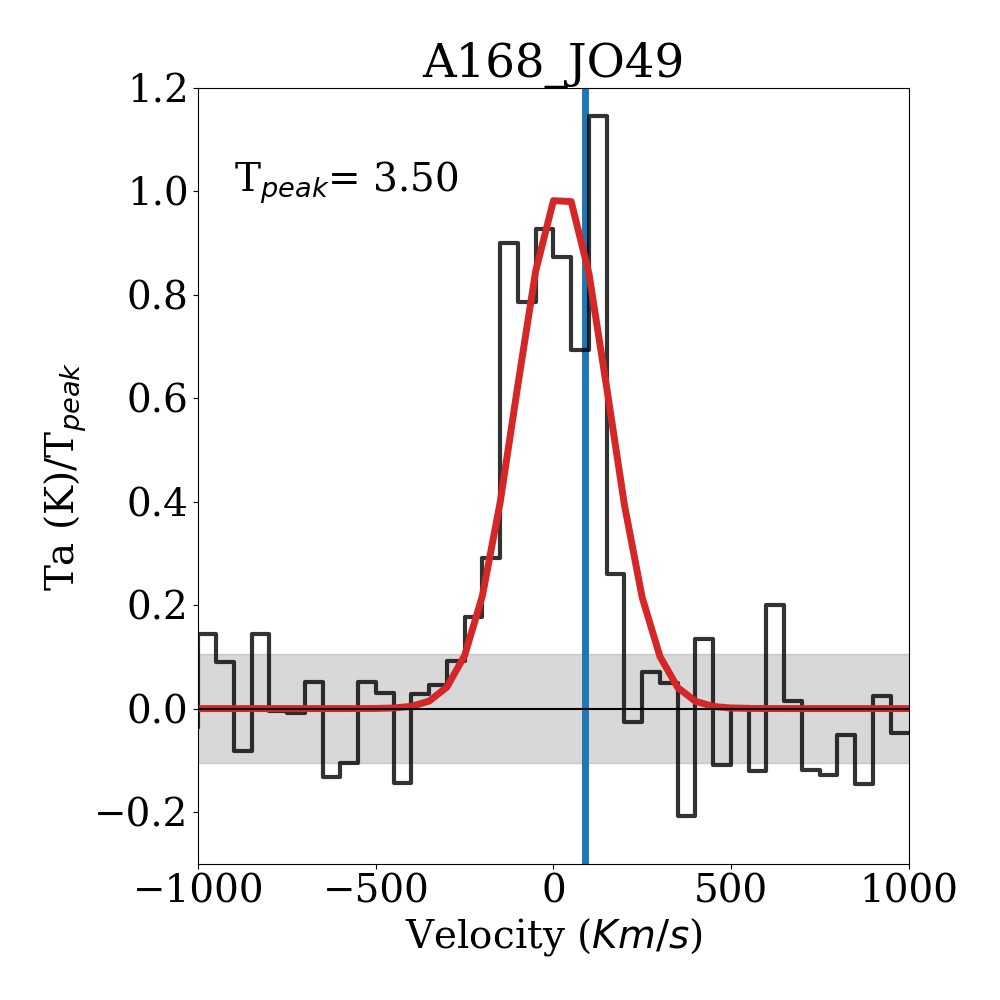}
    \includegraphics[width=0.22\textwidth]{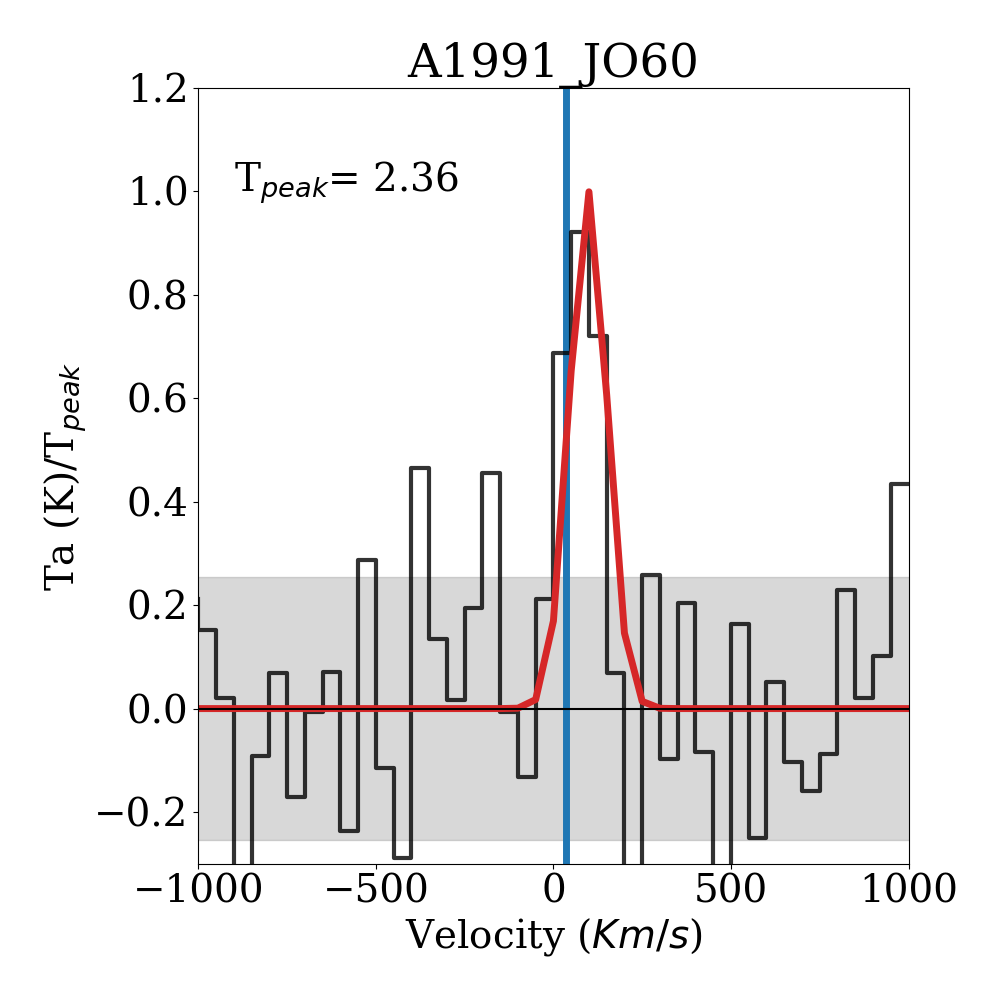}
    \includegraphics[width=0.22\textwidth]{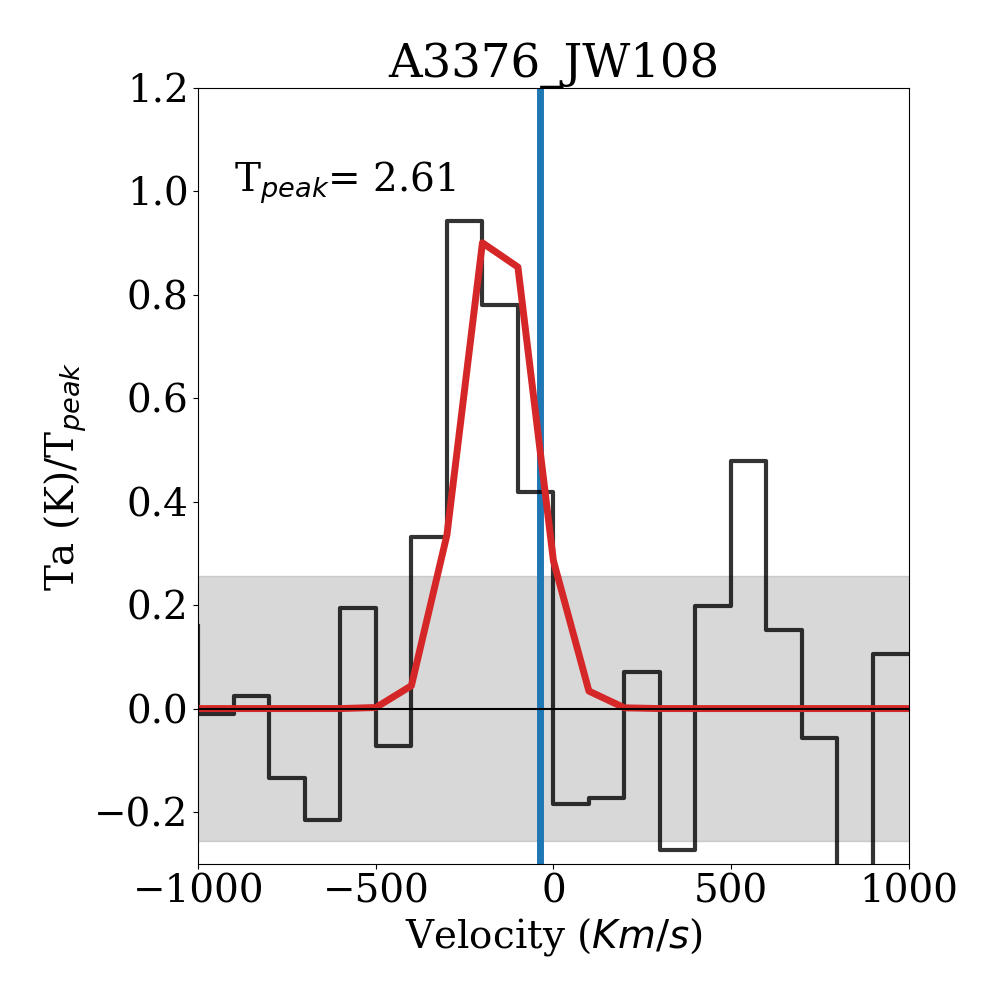}
    \includegraphics[width=0.22\textwidth]{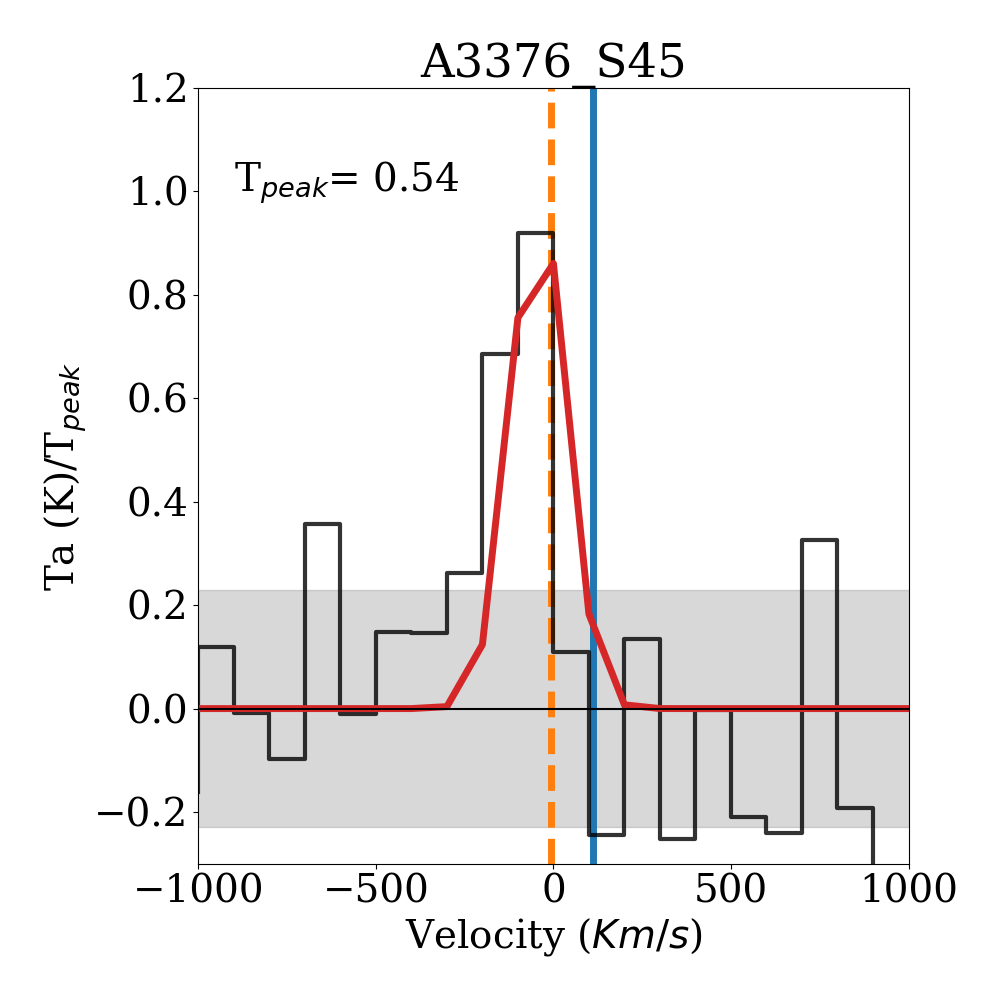}
    \includegraphics[width=0.22\textwidth]{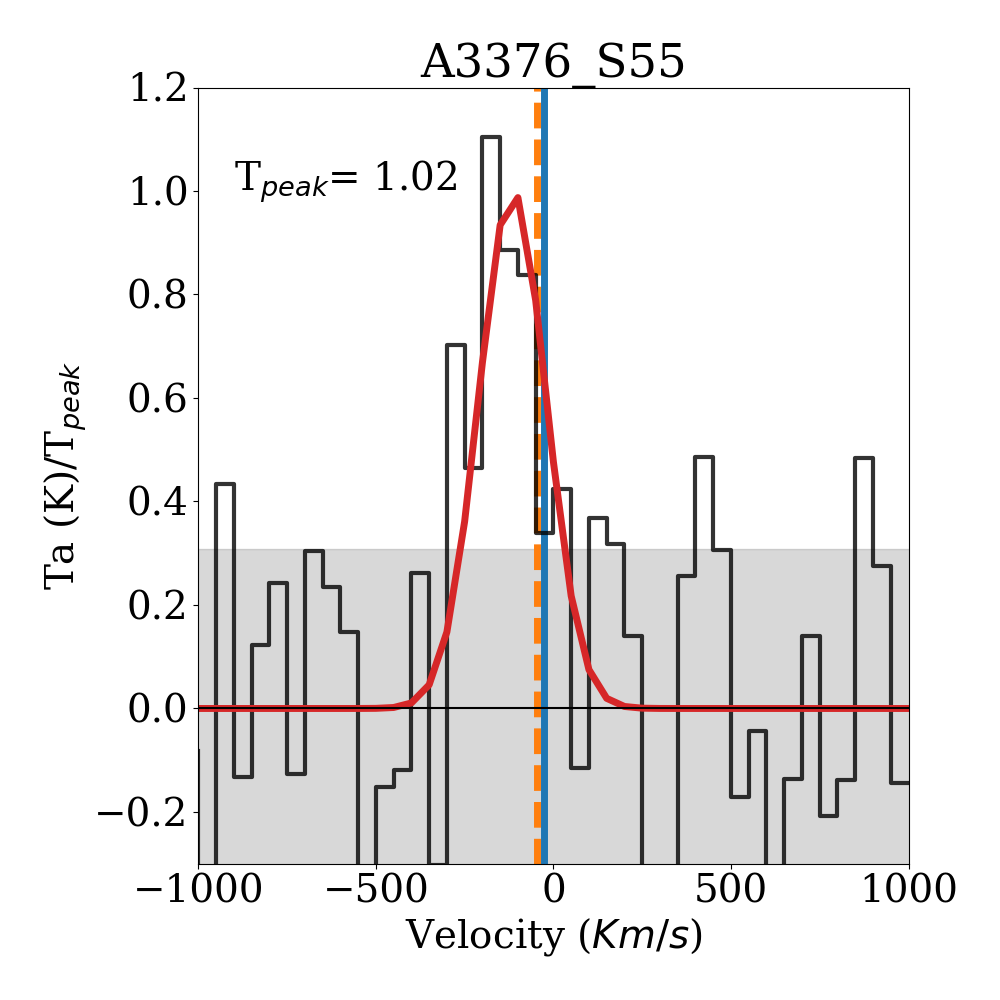}
    \includegraphics[width=0.22\textwidth]{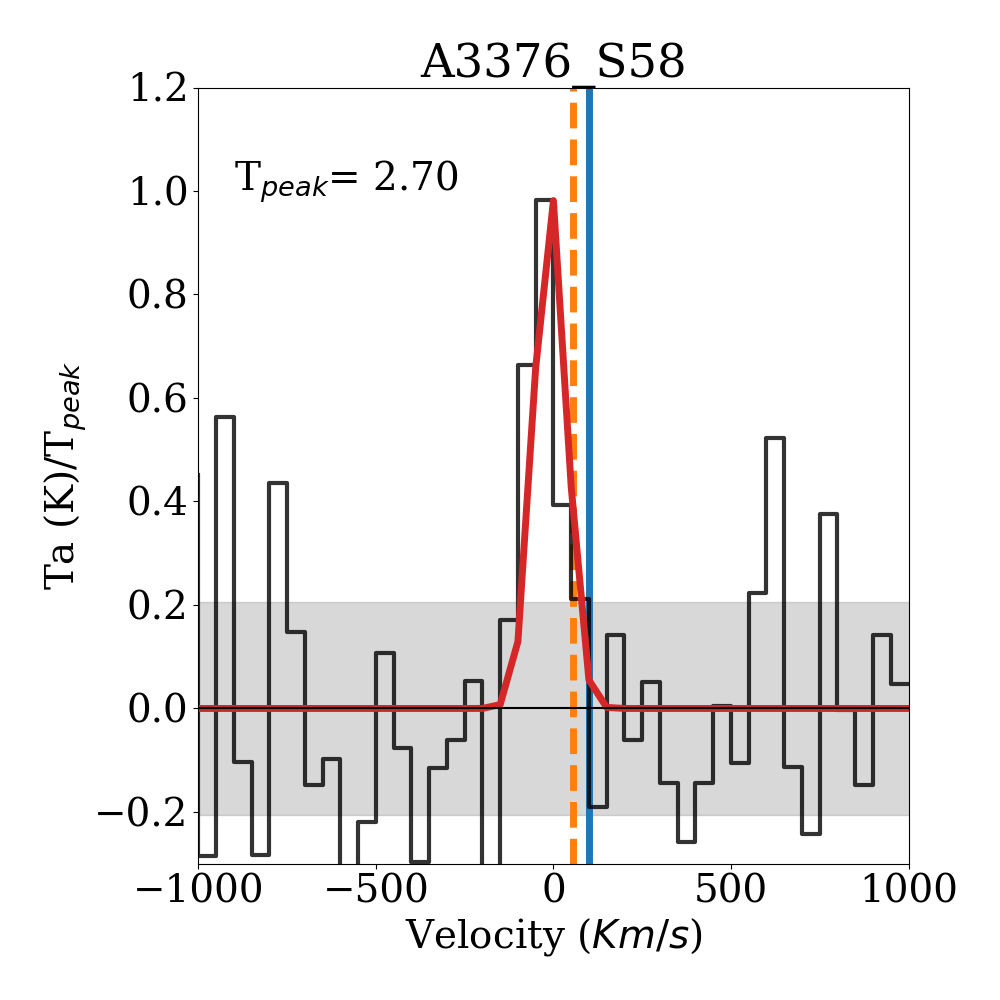}
    \includegraphics[width=0.22\textwidth]{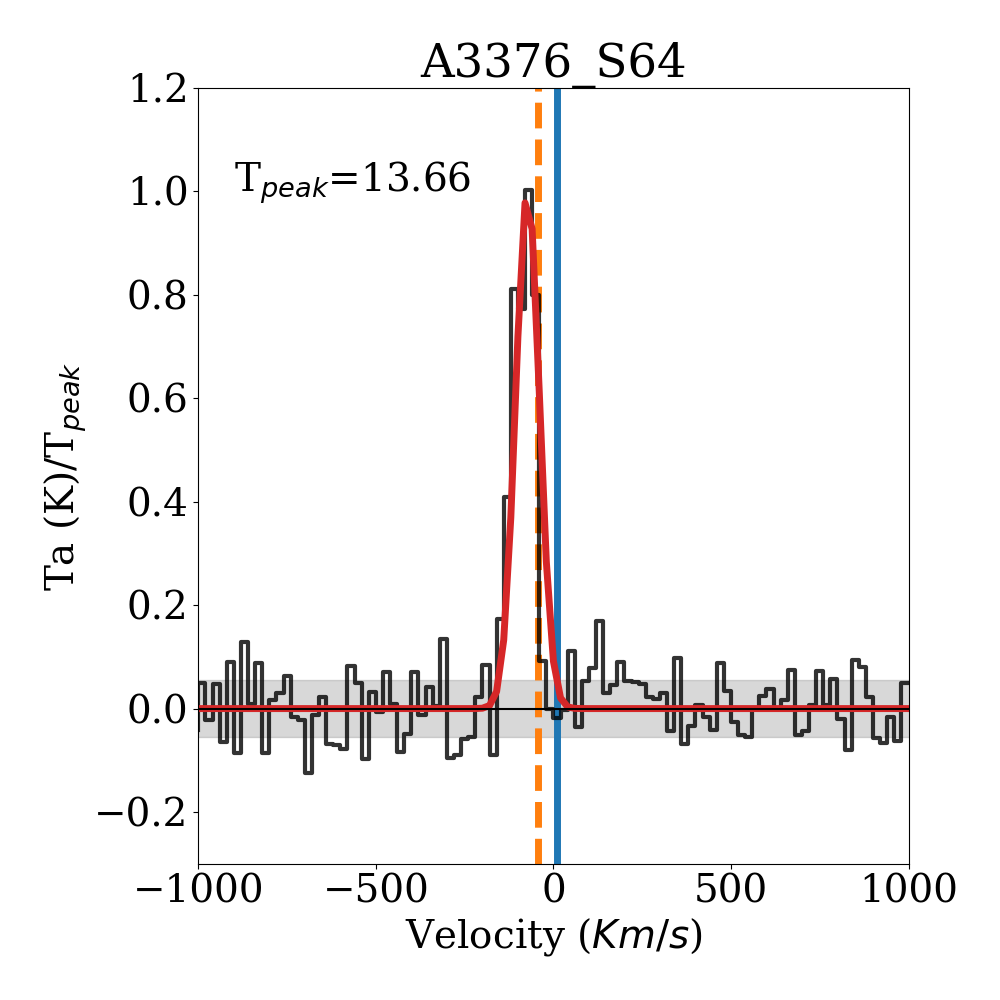}
    \includegraphics[width=0.22\textwidth]{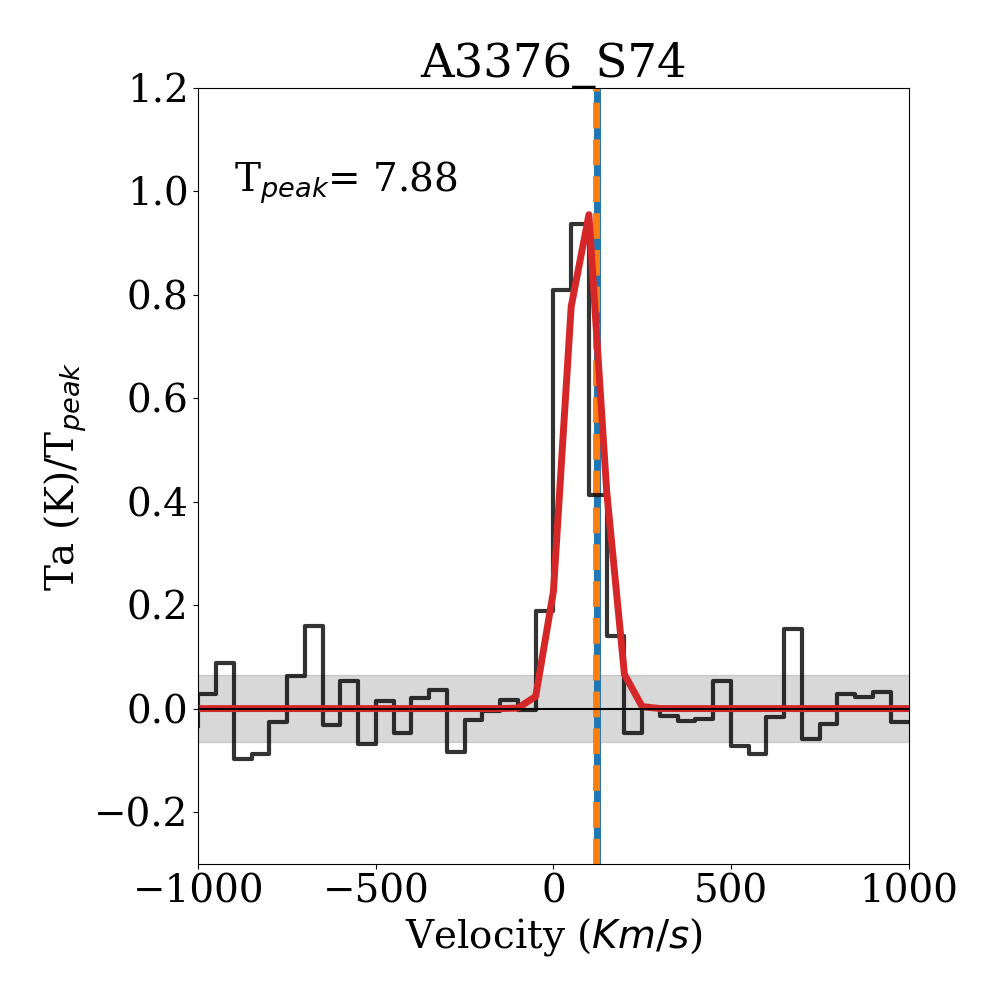}
    \includegraphics[width=0.22\textwidth]{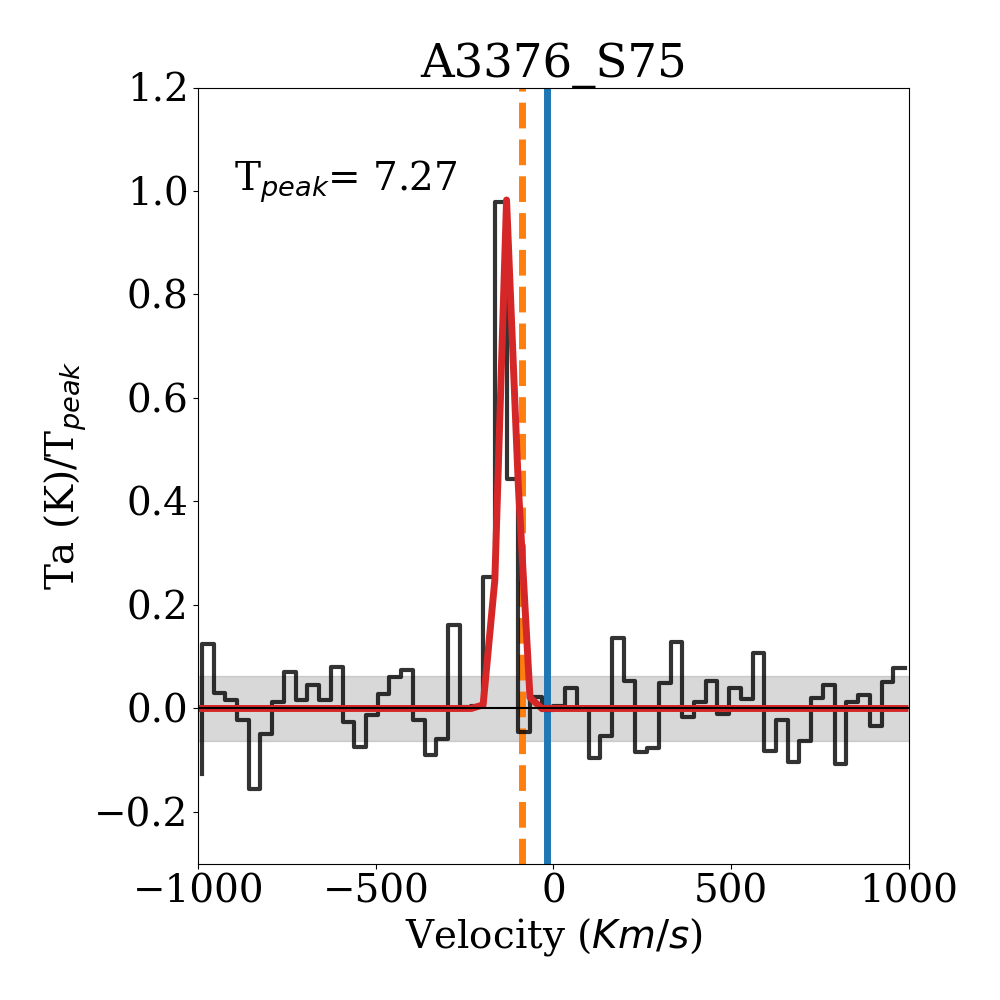}
    \includegraphics[width=0.22\textwidth]{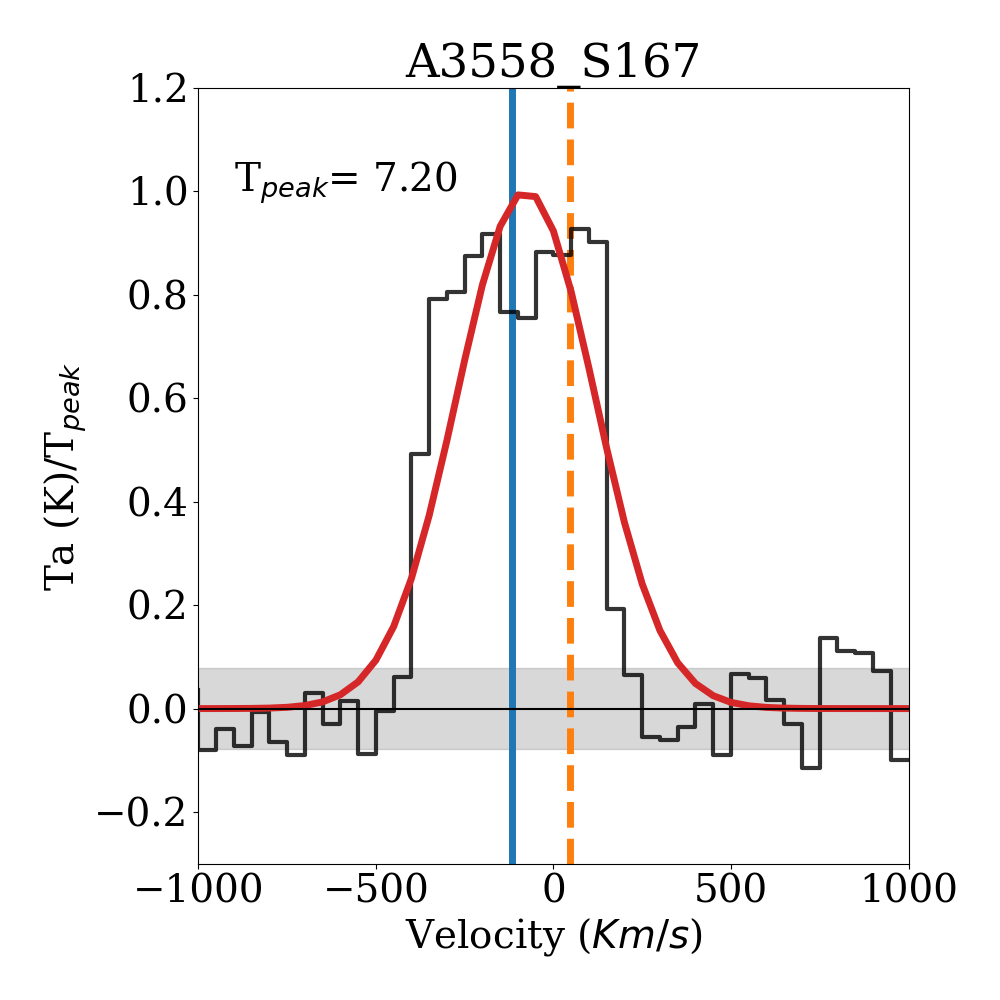}
    \includegraphics[width=0.22\textwidth]{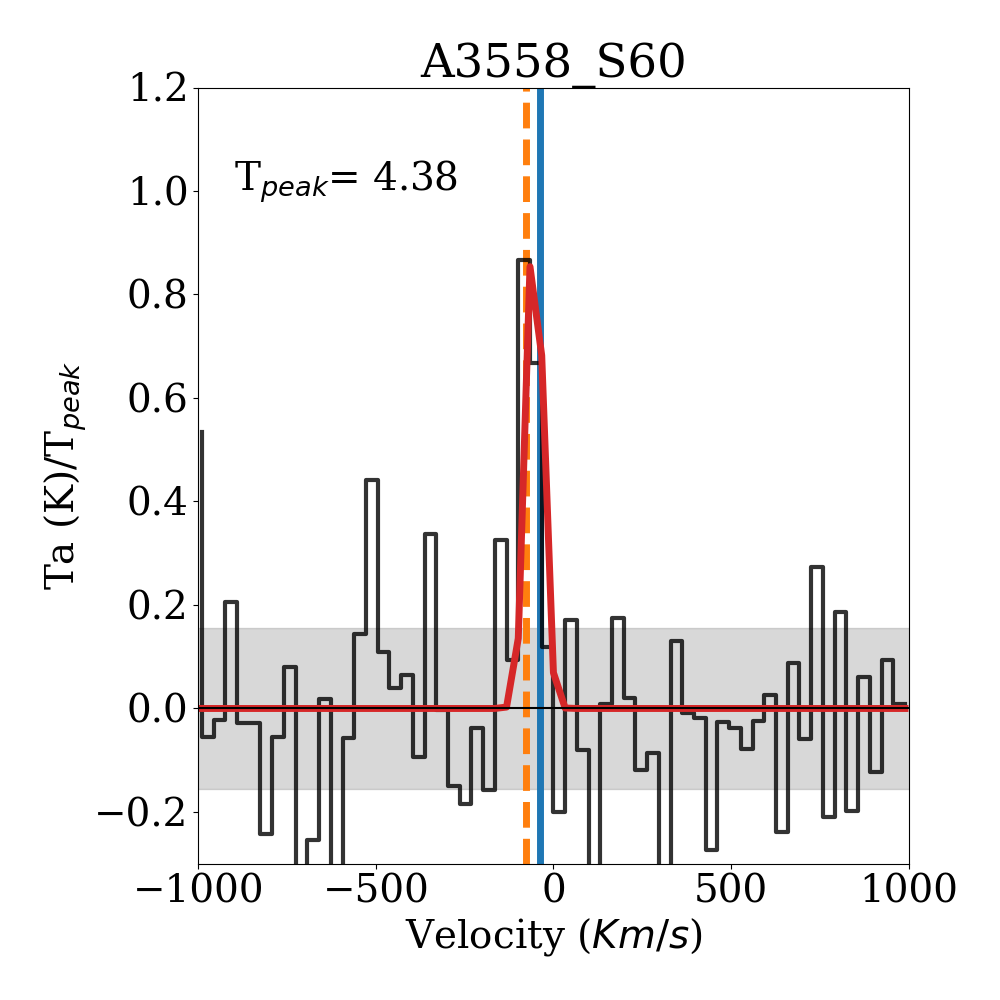}
    \includegraphics[width=0.22\textwidth]{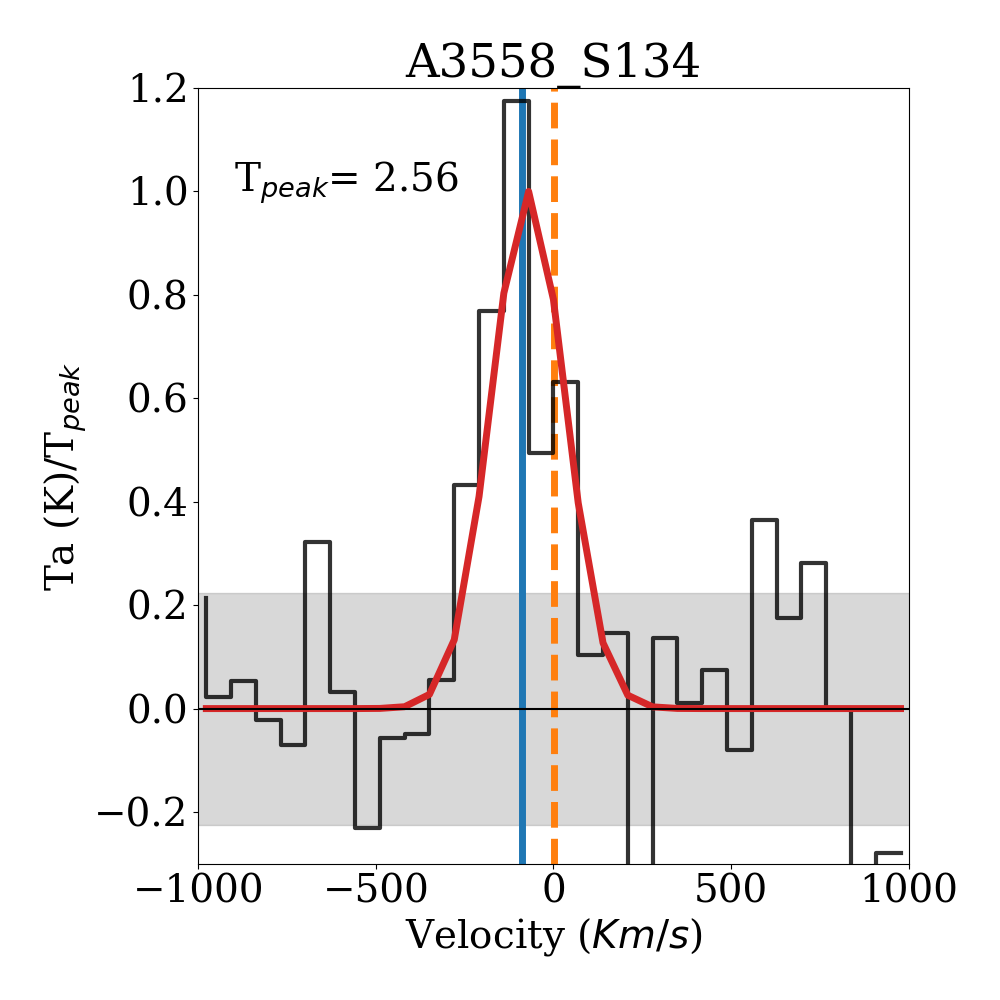}
    \includegraphics[width=0.22\textwidth]{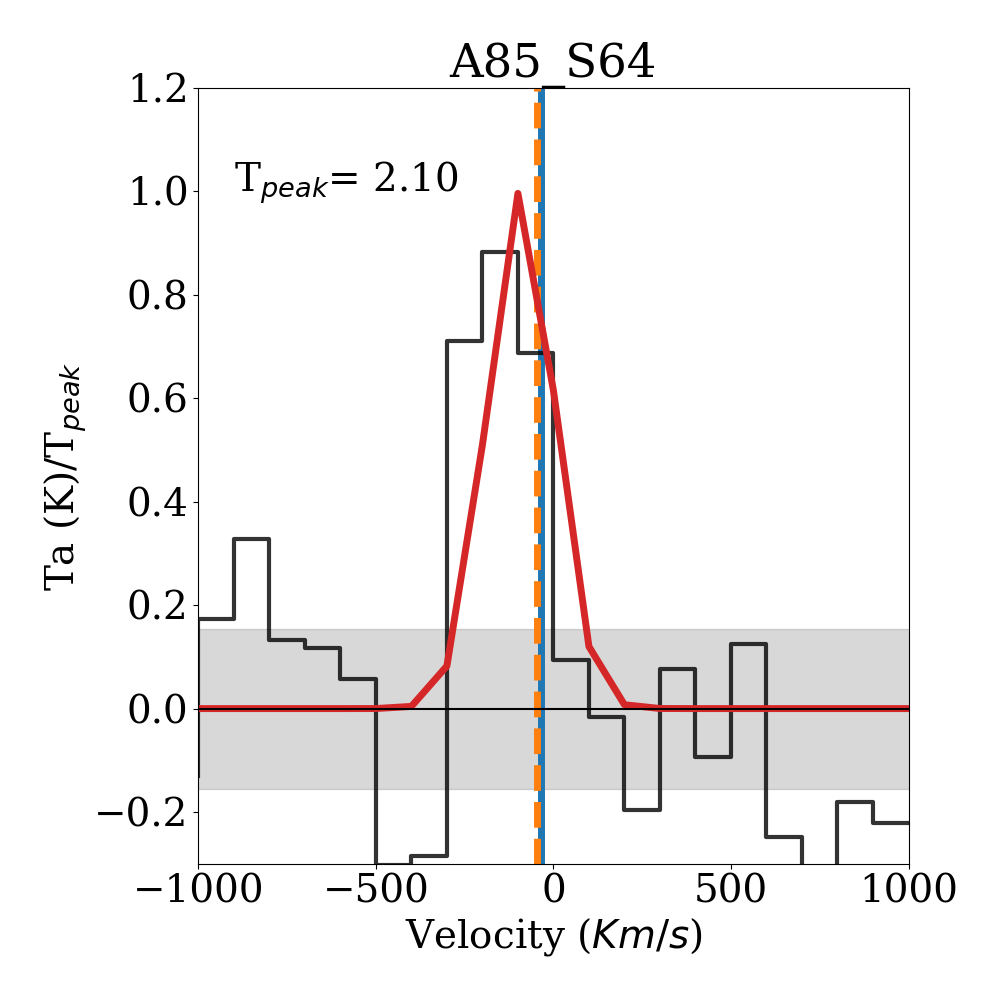}
    \includegraphics[width=0.22\textwidth]{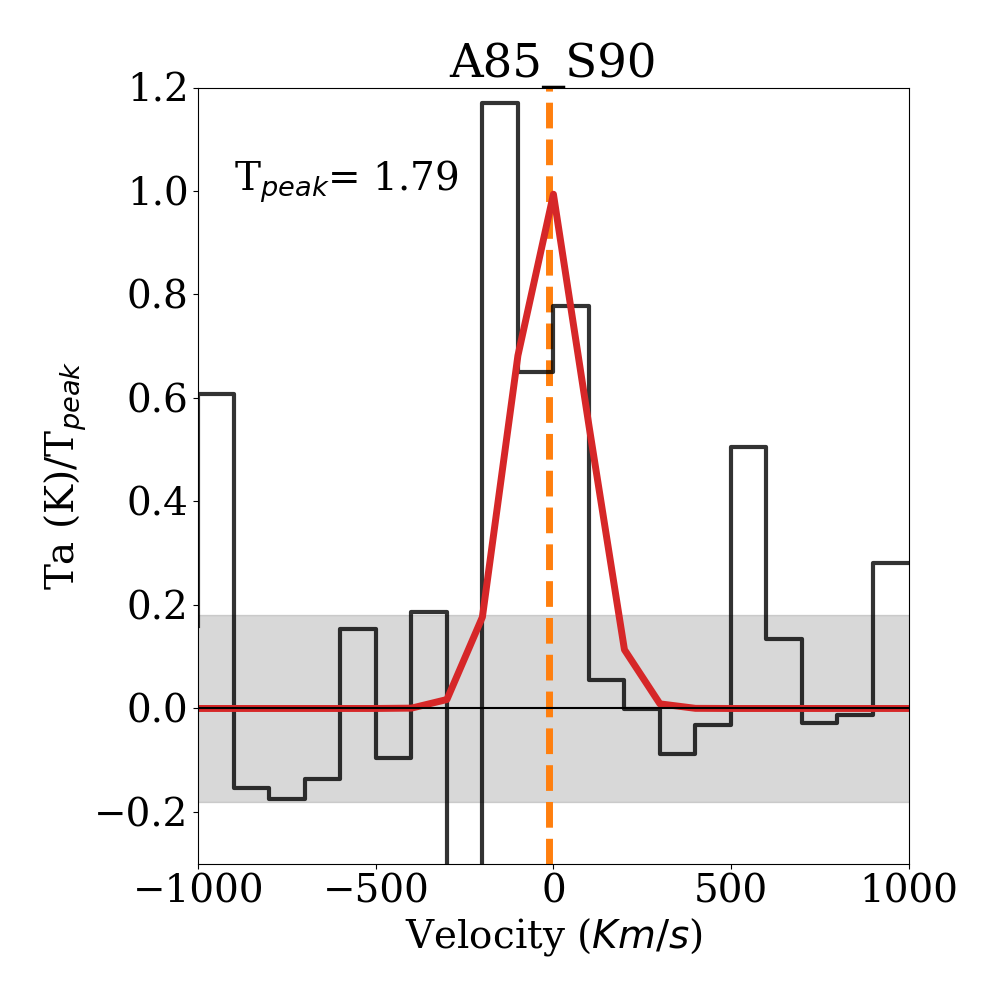}

     \caption{CO(2-1) spectra normalized to the peak temperature for the sources with a S/N$>$3. The red gaussian is the best fit. The blue solid line show the optical velocity while the orange dashed one the \HI velocity}. The grey shaded region is the rms of the observations in the plot range, after excluding the line emission.
    \label{fig:spectra_1}
\end{figure*}

For each source and each spectral resolution we derived 
the peak temperature (main-beam brightness temperature, $\rm T_{mb}$), the linewidth and the position of the CO line relative to the velocity derived from the optical redshift with a single Gaussian fit, as we are here mostly interested in the integrated line flux. Some of our spectra show, though, hints for a double-horn profile, typical of rotating discs.
The extracted CO(2-1) spectra for the central pointing of 14/19 targets with S/N$>$3 are shown in Fig.~\ref{fig:spectra_1}, black lines, with the gaussian fit superimposed in red.
The grey shaded region indicates the rms of the measurement, and has been derived from a 2000 km/s wide spectral range, after excluding the $\pm$1.5 $\sigma$ region centered on the gaussian.

The CO fluxes have been calculated assuming a conversion factor of S$_\nu$/T$_{mb}$ = 39 Jy beam$^{-1}$ K$^{-1}$ for the APEX telescope.
The Precipitable Water Vapour (PWV), the on--source time, and the
results of the fitting procedure of the observations are given in Tab. \ref{tab:results}. When multiple observations were present with significantly different conditions we included in the final dataset the ones that, combined, were giving the highest signal-to-noise and we list in this case the median PWV. The fourth column of the table indicates the velocity resolution (channel width) of the fitted spectrum. 
The errors on the fluxes have been calculated taking into account both the rms of the spectrum and the linewidth (FWHM) determined with the given velocity resolution $\Delta_V$, as
\begin{equation}
    Err_{flux}=\frac{\sigma_{rms} \times FWHM}{\sqrt{FWHM \times \Delta_V}}
\end{equation}
To this term we added a $10$\% error due to the spectral calibration.

One of the nineteen target galaxies (A3376{\_}S84) did not show any significant detection at the expected frequency, and we therefore estimated only an upper limit to its CO flux, and to the corresponding molecular gas mass. We calculated these quantities by using the rms of the observation and a linewidth of 300 km/s, which is the value predicted by the Tully-Fisher relation \citep{Gnedin+2007} for a galaxy with stellar mass of 10$^{11}$ \Msun and seen at 60 deg in the sky-plane.

\begin{table*}
\begin{center}
\caption{Summary of APEX observations: Galaxy ID, median Precipitable Water Vapor of our observations, on source time, RMS of the data, Peak temperature, S/N of the peak, velocity resolution used to measure the CO(2-1) line, Shift in velocity with respect to the galaxy optical redshift, linewidth, CO(2-1) flux, and molecular gas mass.
IDs starting with a t refer to tail pointings.
}\label{tab:results}
\begin{tabular}{|l|c|c|c|c|c|c|c|c|c|c|}
\hline
Galaxy & PWV & T$_{ON}$ & RMS & Tpeak & S/N & Res. & V$_{opt}$-V$_{CO}$   & LW   & Flux & M$_{\rm H_{2}}$  \\
       & mm  & min      & mK  & mK    &     & \kms & \kms                 & \kms & Jy \kms &    1e9 \Msun     \\ 
\hline
A168{\_}JO49    & 0.6  & 30.7  & 0.43 & 3.52 & 8.1   & 33   &  46   & 301 & 44.0   & 5.8$^{+0.7}_{-3.6}$  \\
A1991{\_}JO60   & 2.1  & 30.7  & 0.60 & 2.36 & 3.9   & 50   &  98   & 122 & 12.0   & 2.7$^{+0.4}_{-1.8}$  \\
A3376{\_}JW108  & 3.3  & 17.5  & 0.67 & 2.61 & 3.9   & 100  & -155  & 231 & 25.0   & 3.4$^{+0.5}_{-2.2}$  \\
tA3376{\_}JW108 & 0.9  & 53.7  & 0.18 & 0.33 & 1.8   & 100  & -104  & 195 & 2.7    & 0.4$^{+0.1}_{-0.3}$  \\
A3376{\_}S45    & 1.9  & 145.9 & 0.12 & 0.54 & 4.4   & 100  & -42   & 182 & 4.1    & 0.6$^{+0.1}_{-0.4}$  \\
tA3376{\_}S45   & 2.3  & 130.5 & 0.07 & 0.16 & 2.4   & 150  & -38   & 166 & 1.1    & 0.2$^{+0.1}_{-0.1}$  \\
A3376{\_}S49    & 2.5  & 80.7  & 0.16 & 0.28 & 1.7   & 100  & -288  & 523 & 6.0    & 0.8$^{+0.1}_{-0.5}$  \\
tA3376{\_}S49   & 1.3  & 66.8  & 0.13 & 0.27 & 2.1   & 100  &  3    & 151 & 1.7    & 0.2$^{+0.1}_{-0.1}$  \\
A3376{\_}S55    & 3.6  & 69.0  & 0.31 & 1.02 & 3.2   &  50  & -115  & 223 & 9.4    & 1.3$^{+0.2}_{-0.9}$  \\
A3376{\_}S54    & 3.6  & 69.0  & 0.24 & 0.35 & 1.4   &  50  & -67   & 481 & 7.0    & 1.0$^{+0.2}_{-0.7}$  \\
A3376{\_}S58    & 1.0  & 11.5  & 0.55 & 2.70 & 4.9   &  50  & -9    & 106 & 11.9   & 1.6$^{+0.3}_{-1.1}$  \\
A3376{\_}S64    & 1.0  & 11.5  & 0.76 & 13.7 & 18.1  &  20  & -73   &  79 & 44.6   & 6.1$^{+0.8}_{-3.9}$  \\
tA3376{\_}S64   & 3.9  & 84    & 0.23 & 0.81 & 3.6   & 100  & -83   & 269 & 9.0    & 1.0$^{+0.2}_{-0.8}$  \\
A3376{\_}S74    & 1.0  & 15.3  & 0.51 & 7.89 & 15.6  &  50  & 85    & 116 & 38.0   & 5.2$^{+0.7}_{-3.2}$  \\
A3376{\_}S75    & 1.0  & 30.7  & 0.46 & 7.27 & 15.8  &  33  & -128  &  52 & 15.8   & 2.2$^{+0.3}_{-1.4}$  \\
A3376{\_}S84    & 1.0  & 65.2  & 0.35 & -    & -     &  33  &  -    &  -  & $\leq$4.34   & $\leq$0.6  \\
A3558{\_}S167   & 1.0  & 11.5  & 0.63 & 7.20 & 11.5  &  50  &  -77  & 457 & 136.6  & 20.7$^{+2.3}_{-12.7}$ \\
tA3558{\_}S167  & 1.0  & 38.4  & 0.29 & 1.03 & 3.6   &  50  &  75   & 134 & 5.7    & 0.9$^{+0.1}_{-0.6}$  \\
A3558{\_}S60    & 1.0  & 11.5  & 0.66 & 4.38 & 6.6   &  33  &  -53  & 54  & 9.8    & 1.5$^{+0.3}_{-1.0}$  \\
tA3558{\_}S60   & 1.0  & 25.5  & 0.42 & 1.13 & 2.7   &  33  & -97   & 41  & 1.9    & 0.3$^{+0.1}_{-0.2}$  \\
A3558{\_}S124   & 1.0  & 19.2  & 0.53 & 0.75 & 1.4   &  33  &  -61  & 82  & 2.6    & 0.4$^{+0.2}_{-0.3}$  \\
A3558{\_}S134   & 1.0  & 8.9   & 0.51 & 2.56 & 5.0   &  70  &  -71  & 244 & 26.0   & 3.9$^{+0.6}_{-2.5}$  \\
tA3558{\_}S134  & 1.0  & 58.4  & 0.16 & 0.25 & 1.6   & 150  &  73   & 94  & 0.97   & 0.1$^{+0.1}_{-0.1}$  \\
A85{\_}S57      & 0.4  & 23.1  & 0.18 & 0.35 & 1.9   & 150  &  27   & 289 & 4.2    & 0.8$^{+0.2}_{-0.5}$  \\
A85{\_}S64      & 0.2  & 7.7   & 0.43 & 2.10 & 4.9   & 100  &  -92  & 219 & 19.1   & 3.9$^{+0.5}_{-2.5}$  \\
tA85{\_}S64     & 0.2  & 27.0  & 0.36 & 1.27 & 3.5   &  33  &  35   & 89  & 4.7    & 0.9$^{+0.2}_{-0.7}$  \\
A85{\_}S90      & 1.2  & 13.4  & 0.33 & 1.79 & 5.4   & 100  & -11   & 239 & 17.7   & 3.6$^{+0.4}_{-2.3}$  \\
\hline
\end{tabular}
\end{center}
\end{table*}

Among the other 18 galaxies 4 have an emission detected with a low S/N ($<3$).

The derived CO velocities of the target galaxies are very close to those inferred from the optical redshift, indicated in Fig.~\ref{fig:spectra_1} by a blue vertical solid line. The difference between the two velocities are indicated in Col. 8 of Tab.\ref{tab:results}, and are mostly below the velocity resolution.
For the galaxies having a MeerKAT detection we also show the \HI velocity as a dashed line, which is often closer to the molecular gas velocity with respect to the optical stellar one. The largest differences in systemic velocity are found in galaxies that appear nearly face-on.
For the massive galaxy A3558{\_}S167 (GASP JO147) 
the double horn profile indicates that the disk rotation is resolved in velocity.
All the galaxies that have a GASP counterpart (i.e. signs of stripped \Ha-emitting gas) except the low mass A3558{\_}S124 (JO159), have a significant emission of molecular gas, with S/N ranging from $\sim$4 to $\sim$16. 
When a signal was detected, the observations reached the desired S/N level using a significantly lower integration time, with respect to what we could predict assuming that cluster galaxies were \hdue poor, i.e. having a molecular gas content  3$\sigma$ below the average content of galaxies with a similar stellar mass.
This immediately confirms that in most cases our \HI/optical selected sample of candidate ram-pressure stripped galaxies are not \hdue deficient.

\begin{figure}
    \centering
    \includegraphics[width=0.22\textwidth]{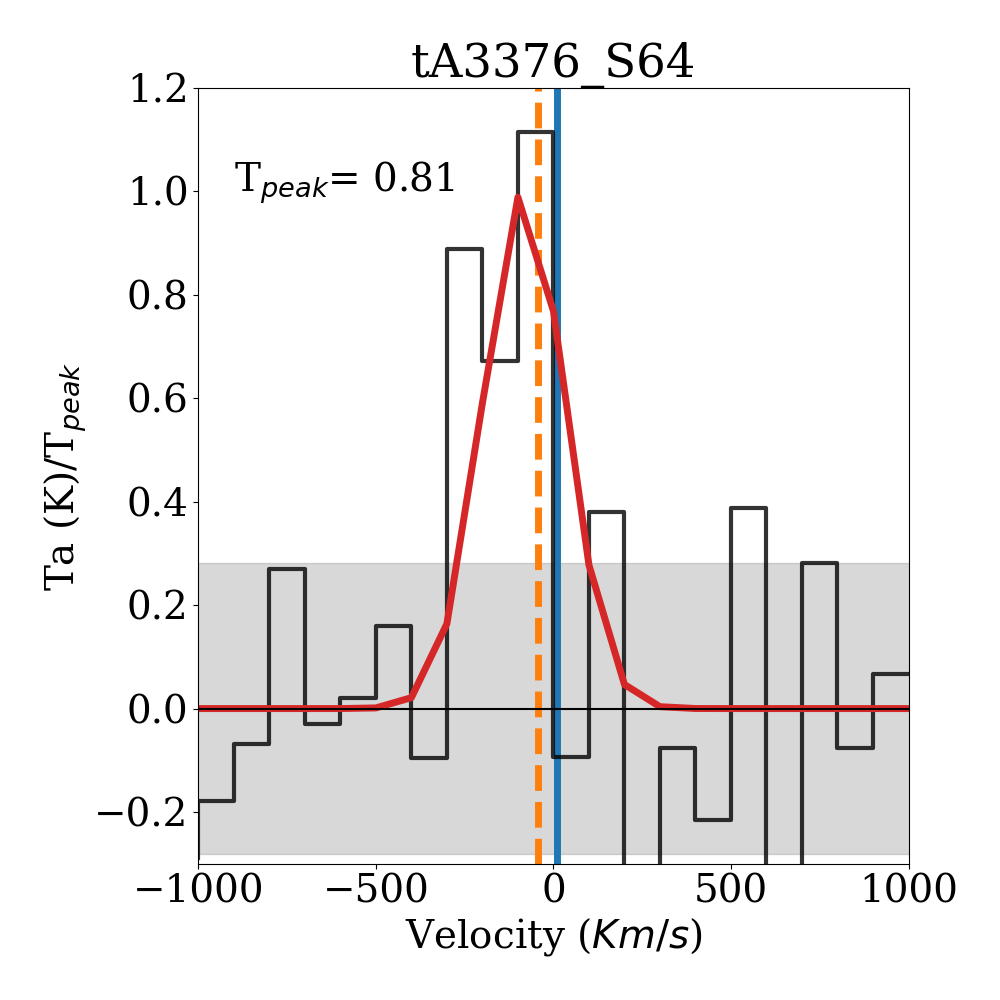}
    \includegraphics[width=0.22\textwidth]{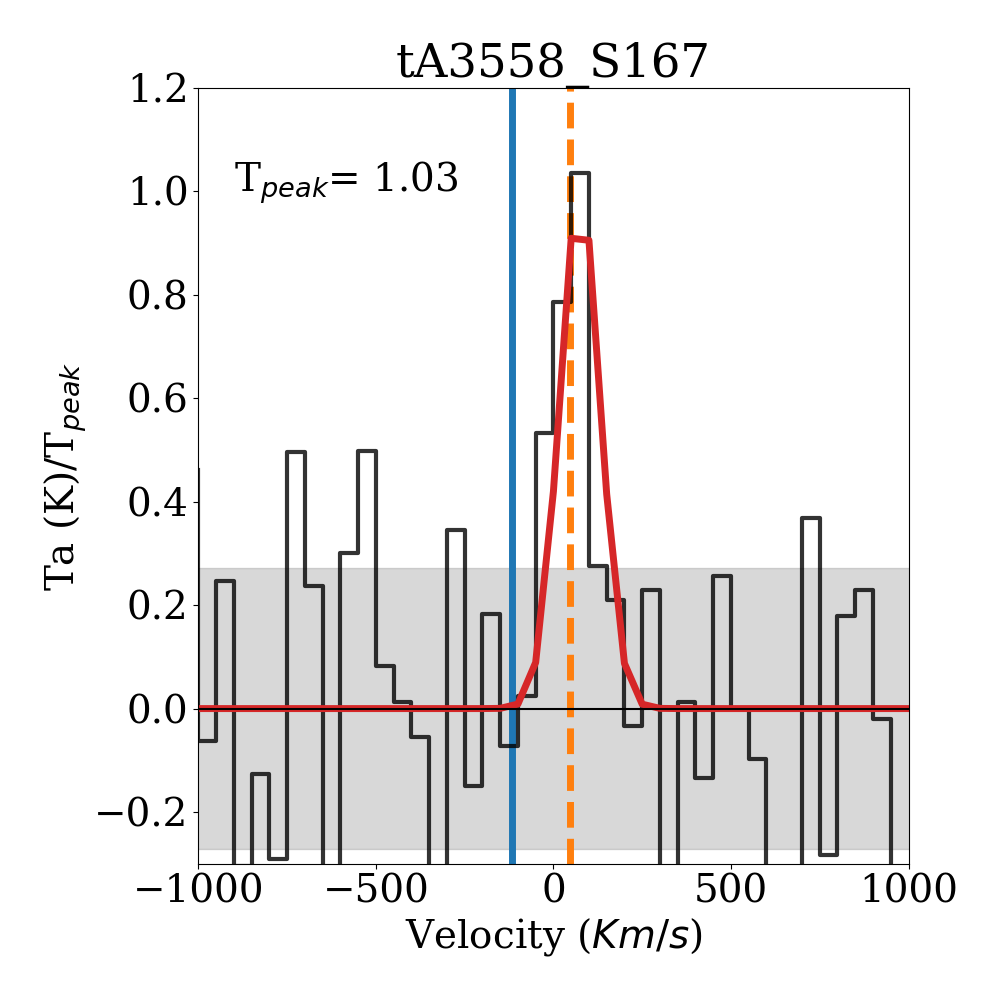}
    \includegraphics[width=0.22\textwidth]{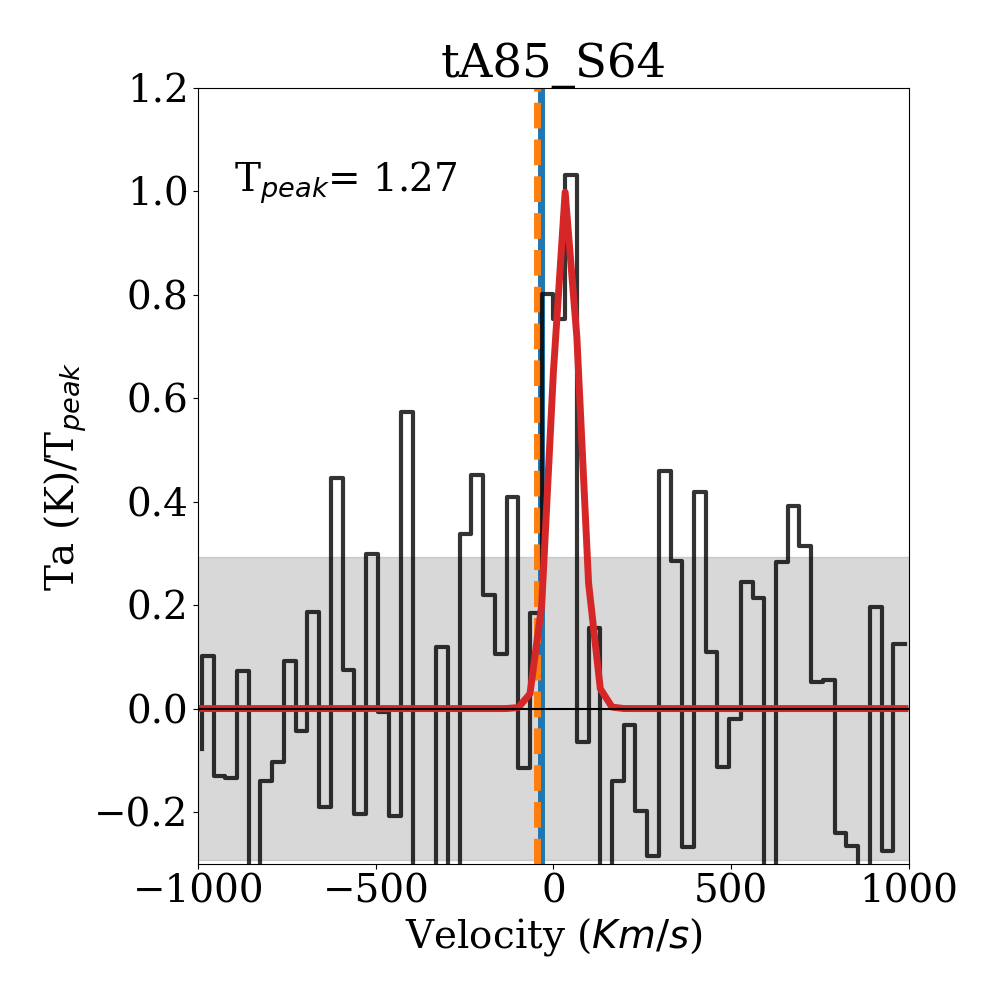}
    \includegraphics[width=0.22\textwidth]{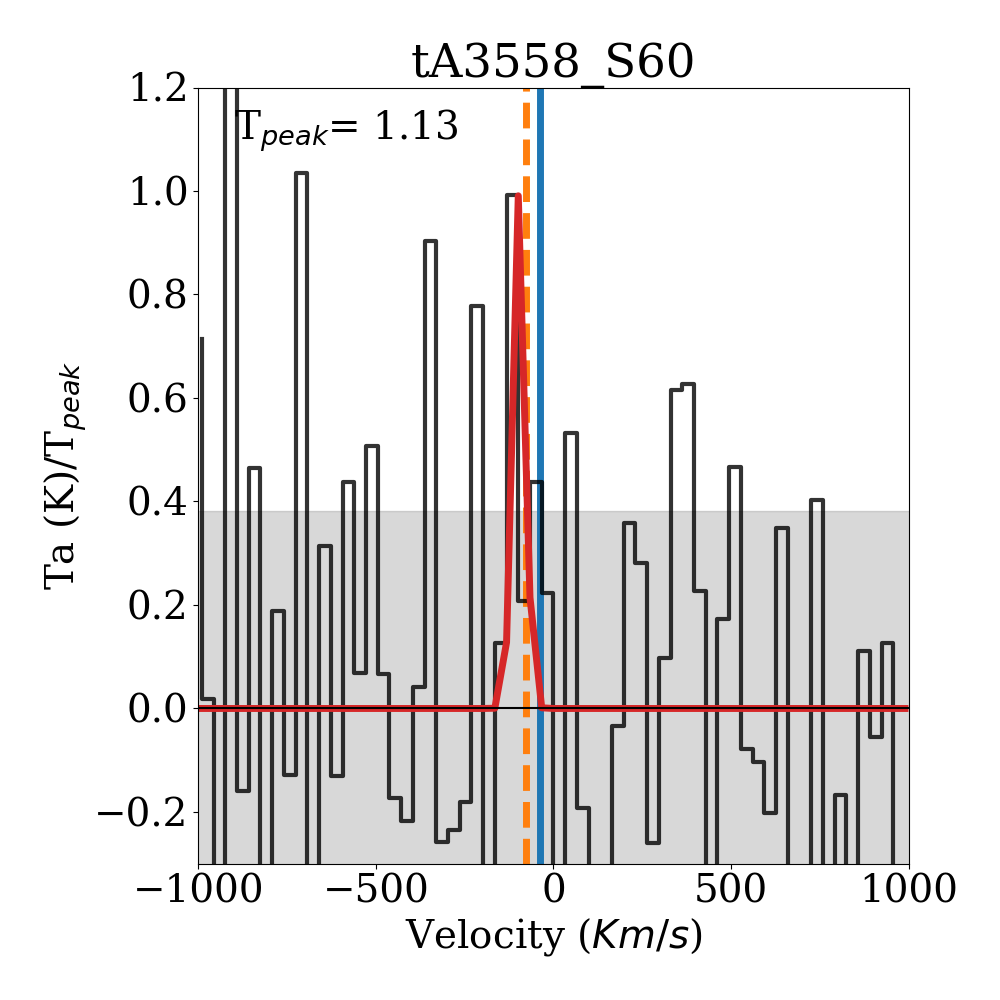}
     \caption{CO(2-1) spectra normalized to the peak temperature for the tail observations with a S/N$>$3 (excluding the marginal emission shown in the lower right corner which has S/N=2.7). The red gaussian is the best fit. The blue solid line show the optical velocity while the orange dashed one the \HI velocity}. The grey shaded region is the rms of the observations in the plot range, after excluding the line emission.
    \label{fig:spectra_2}
\end{figure}

For the APEX pointings centered on the supposed gas tails identified either following the \HI extension or the \Ha tail, we obtained a significant detection in 3/8 pointings, namely  A3376{\_}S64, A3558{\_}S167 (JO147) and A85{\_}S64 (JO200), shown in Fig.~\ref{fig:spectra_2}, together with the marginal emission in the tail of A3558{\_}S60 (JO157), which has S/N=2.7.
All the other tail pointings have a low S/N (from 1.6 to 2.4) and are characterized by a relatively narrow linewidth.

\section{The molecular gas content}\label{sec:gascontent}

In order to be consistent with our previous works \citep{GASPX,Moretti_2020}, we used the measured line fluxes to calculate then the \hdue mass using the following equation from \citet{WatsonKoda2016}:

\begin{equation}\label{eqn:wk_co21}
\left(\frac{M_{\rm H_2}}{M_{\odot}}\right) = 
3.8 \times 10^3 \left( \frac{\alpha_{10}}{4.3}\right)
\left(\frac{r_{21}}{0.7}\right)^{-1}
\left(\int S_{21}dv \right)
\left(D_L \right)^2
\end{equation}
where 
$\rm \alpha_{10}$ is the CO-to-$\rm H_2$ conversion factor expressed in $M_{\odot}pc^{-2}$(K km s$^{-1}$)$^{-1}$  \citep{Bolatto2013}, $r_{21}$ is the CO(2-1)-to-CO(1-0) flux ratio, $S_{21}$ is the CO integrated line flux in Jy and D$_L$ is the luminosity distance in Mpc.
As done in \citet{GASPX} we used $\rm \alpha_{10}=4.3$, i.e. the standard Milky Way value corresponding to a conversion factor of $2\times 10^{20}$ cm$^{-2}$(K km s$^{-1}$)$^{-1}$ including the helium correction. As for the $r_{21}$, we adopted the same value (0.79) that we used in our APEX \citep{GASPX} and ALMA \citep{GASPXXII,Moretti_2020} data analysis for the sake of comparison.
To take into account the systematic uncertainty on the \aco factor, we calculated a lower limit on the mass obtained by assuming that its value is 50\% lower than the canonical Milky Way factor, while higher values are less probable in massive, metal-rich galaxies \citep{Bolatto2013}. Errors calculated in this way are shown in the last column of Tab.\ref{tab:results}.
The molecular gas content of the sample galaxies varies between 10$^9$ and 2$\times$10$^{10}$ \Msun in galaxies where the line is measured with sufficient S/N. 

The molecular gas masses calculated in this way are systematically 5\% larger than those calculated using the \cite{Solomon2005} equation, which explicitly takes into account the dependence on redshift, being
\begin{equation}\label{eqn:svb05}
\left(\frac{M_{\rm H_2}}{M_{\odot}}\right) = 
\alpha \times L'_{CO}=
\alpha \times 3.25 \times 10^7 S_{CO}\Delta v\nu_{obs}^{-2} \left(D_L \right)^2 \left(1+z \right)^{-3}
\end{equation}
where $S_{CO}\Delta v$ is the velocity integrated flux in Jy \kms, $\nu_{obs}$ is the observed line frequency at the source redshift z, and $D_L$ is the corresponding luminosity distance.
In this formulation the major source of uncertainty comes from the $\alpha$ conversion value, which includes both the CO-to-$\rm H_2$ conversion factor and the $r_{21}$ line ratio, and is typically assumed to be equal to 5.5 \citep{Jachym2014}.
These values are, in any case, within the errors of our measurements.

This allows a straightforward comparison with the recent determination of the molecular gas content/fraction in Virgo galaxies by \cite{Brown+2021}, who also uses the \cite{Chabrier2003} IMF to determine galaxy stellar masses.

\begin{figure*}
    \centering
    \includegraphics[width=0.95\textwidth]{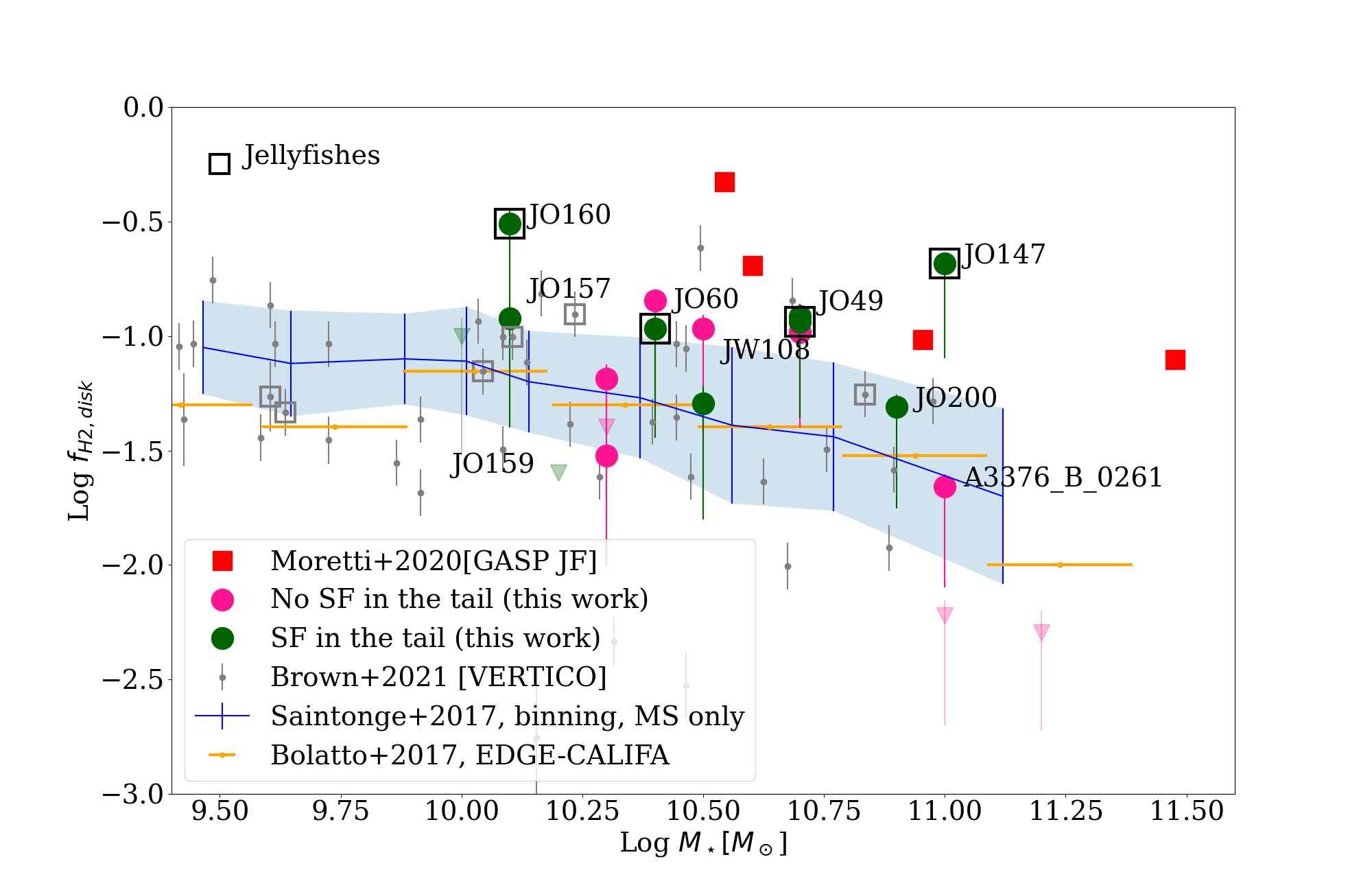}
     \caption{Molecular gas fraction $f_{H_2}(=M_{H_2}/M_{\star}$) as a function of the galaxy stellar mass: the blue line shows the mean scaling relation found in the xCOLDGASS survey by \citet{Saintonge2017} for field galaxies on the main sequence with its dispersion as blue shaded region, while orange lines refer to spiral galaxies from the EDGE-CALIFA survey \citep{Bolatto+2017}.
     Grey dots are Virgo galaxies from \cite{Brown+2021}. Red squares are GASP jellyfish galaxies from \cite{Moretti_2020}, with measured ALMA fluxes.
 Green and magenta symbols show the APEX measurements for galaxies with and without a star forming tail, with transparent triangles showing the low S/N detections (and the upper limit for the A3376{\_}S84 source). Symbols enclosed in squares are jellyfish galaxies, either from GASP (black) or Vertico (grey). Galaxies having a MUSE counterpart (Tab. \ref{tab:targets_data}) are labeled with the GASP ID.}
    \label{fig:fh2}
\end{figure*}
Fig.~\ref{fig:fh2} shows the position of our targets in the molecular gas fraction versus stellar mass plane, where we indicated with different colors galaxies with (green) and without (magenta) star formation in the tail. The red squares refer to the 4 GASP galaxies undergoing strong RPS analyzed in \cite{Moretti_2020}. 
It can be immediately seen that $\sim 50$ \% of the observed sample of \HI selected galaxies (9/19) with asymmetries lie above the relation found for normal star forming galaxies in the field (blue dashed region from \citealt{Saintonge2017}) and in Virgo (grey dots).
Among the galaxies closer to the relation for field galaxies, A85{\_}S64 (JO200) possesses a MUSE \Ha counterpart, which follows its spiral arms.
We covered this emission with a second APEX pointing, which reveals the presence of molecular gas. 
The \Ha morphology of this galaxy closely resembles the one of the JO201 GASP galaxy \citep{GASPII}, which shows unwinding spiral arms possibly caused by the RPS \citep{GASPXXIX}.
 If we include the molecular gas mass in the tail in the global budget, then the point would move even closer to the upper edge of the distribution of field galaxies in the M$_{\star}$-f$_{H_2}$ plane (see Fig.~\ref{fig:fh2}).

Among the galaxies located below the relation in Fig.~\ref{fig:fh2}, only one has a clear detection (A3376{\_}S45) indicating that this galaxy is \hdue deficient. This galaxy also shows no \hdue emission from the tail region. 
The other galaxies located below the \citet{Saintonge2017} relation are characterized by a low signal-to-noise (shaded triangles), irrespective of their stellar mass.
One of these, the GASP galaxy JO159, has been classified as an initial stripping phase (Poggianti+, in prep.).

It is clear that when the \HI asymmetry is coupled with having SF in the tail, then in 75\% (6/8) of the cases with S/N$>$3  the \hdue content is enhanced within the stellar body (83\% = 10/12 if we include the \citealt{Moretti_2020} galaxies).
In particular, galaxies classified by GASP as jellyfishes (red squares and green dots surrounded by a black square in Fig.~\ref{fig:fh2}), i.e. those having an ionized gas tail at least as long as the stellar disk and therefore being at their peak stripping phase, show a strongly enhanced molecular gas fraction.
Among these, A3558{\_}S167(GASP JO147) is the \hdue-richest massive galaxy (see also \citealt{Merluzzi2013}) and, together with the galaxy JO49, also \hdue rich, it hosts an active nucleus, as the other massive jellyfishes in GASP \citep{poggianti2017,Peluso+2021}.
Noticeably, also the GASP galaxy JW108, which has been classified as a truncated disk, i. e. a galaxy where the ionized gas emission comes from a region smaller than the extent of the stellar disk, typical of the latest stage of gas stripping, lies above the relation.
In fact, because of the more advanced stage of stripping of this galaxy, most of the \HI gas is probably removed or got converted into \hdue, resulting in a very faint \HI emission (below the detection threshold) in this galaxy, while still showing an increased amount of cold molecular gas in its disk (while the detection in the tail has a low S/N).
In fact, besides JO159, the only GASP galaxy not lying above the field relation is A3376{\_}S75 (A3376{\_}B{\_}0261) which is a control sample galaxy in GASP and thus shows a pretty normal molecular gas content.

On the other hand, when the \HI asymmetry manifests in galaxies without SF in the tail the \hdue content within the stellar body tend to be normal/low in 50\%  (3/6) of the cases  (magenta dots in Fig.~\ref{fig:fh2}, representing detections with S/N$>$3). If we include the low S/N detections, this percentage increases to 67\% (6/9). 

Our result seems to suggest that the different gas proportion and displacement follow a temporal evolution: first the galaxy is hit by the ram pressure and starts developing the \HI asymmetry, possibly losing molecular gas as well (at least the more diffuse component). At the same time, in the regions of the disc compressed by ram pressure the \HI is  efficiently converted in molecular gas in the disk, and by the time we see the ionized gas tail the molecular gas fraction of the galaxy is significantly enhanced. 
This increase in the molecular gas fraction seems to be preserved also in galaxies in advanced stripping stage, as is the case of A3376{\_}JW108. 
This result is perfectly in agreement with the detection of large cold gas reservoir in post-starburst (PSB) galaxies  \citep{French+2015,Rowlands+2015,Alatalo+2016}, under the hypothesis that these objects belong to the evolutionary sequence caused by the RPS \citep{GASPXXIV,Werle+2022}. In fact, in PSB galaxies the cold gas is 
thought to be present in a diffuse phase not contributing to the star formation.

\subsection{Molecular gas in the stripped tails}

The molecular gas content measured in the pointings covering the expected gaseous tail amount to $\sim$10$^{9}$ \Msun in all the three sources with S/N$>$3, namely A3376{\_}S64, A3558{\_}S167 (JO147) and A85{\_}S64 (JO200).
Interestingly, two of these tails  belong to very massive GASP galaxies (JO200 and JO147) and exhibit a narrow CO linewidth ($\sim$90 and $\sim$130 \kms, respectively), while the third one is a low-mass galaxy showing a broad linewidth.
All the low S/N tail detections, instead, are related to galaxies with low stellar masses (logM$_\star$=10.1-10.5 \Msun ).
This result suggests that in our sample, when the ram-pressure is able to produce a long gas tail visible at optical/UV wavelengths, a cold molecular gas counterpart is detectable.
In low mass galaxies the amount of cold gas in the tail is proportionally lower implying a more difficult detection.

The nFLASH APEX detector, though, can only observe the CO(2-1) line emission, usually associated with the dense gas, while the less dense phase could be more easily stripped and found along the tails.
Unfortunately, APEX can not observe this transition, and further data (e.g. ALMA) are needed to evaluate the diffuse molecular gas content in ram-pressure stripped tails.

\section{Cold gas balance}\label{sec:coldgasratio}

For the majority (16/19) of the galaxies having a molecular gas mass determination from APEX, we also have the \HI cold gas content, as derived from MeerKAT data\footnote{Two galaxies in the sample are not covered by MeerKAT observations (A169\_JO49 and A1991\_JO60)}.
As already shown in \cite{Moretti+2022}, in case a galaxy is massive and at its peak stripping, such as the four ones observed with ALMA, its cold gas fraction \hdue/HI is significantly enhanced with respect to normally star forming field galaxies not subject to the cluster dense environment and/or ram pressure stripping.
More recently, \cite{Zabel+2022} compared the atomic and molecular gas content in Virgo galaxies, which were divided into different classes based on their \HI morphology and content, following the classification criteria proposed by \cite{Yoon+2017}. 
\citet{Zabel+2022} found a weak correlation between the \HI and H$_2$ deficiencies, albeit with a large scatter, suggesting that the \HI deficiency does not always predict H$_2$ deficiency. 
This was interpreted as an indication that ram pressure stripping is not effective at reducing global molecular gas fractions on the timescales in which such features are still clearly visible.

 With the aim of comparing our galaxies with those in Virgo, we used both the visual asymmetries and the cold gas content (\HI and \hdue) to classify our galaxies using the definitions in \citet{Yoon+2017} (classification given in the last column of Tab.\ref{tab:targets_data}).
More in detail, we calculated the \HI and \hdue deficiencies using the definition based on the stellar mass \citep{Zabel+2022}
\begin{equation}
    \mathrm{def}_i = \log M_{i,\mathrm{exp}} - \log M_{i,\mathrm{obs}} \, ,
\end{equation}
where $i=$HI and \hdue, $\log M_{i,\mathrm{exp}}$ is the \HI or \hdue mass expected given the galaxy stellar mass, and $\log M_{i,\mathrm{obs}}$ is the observed \HI or \hdue mass. 
In particular, $M_{i,\mathrm{exp}}$ is the median \HI (\hdue) mass calculated in bins of stellar mass using the xGASS (xCOLD-GASS) control sample \citep{Saintonge2017,Catinella+2018}. 
We used ten bins of stellar mass equally spaced in logarithmic scale. 

\begin{figure}
	\centering
	\includegraphics[width=1\columnwidth]{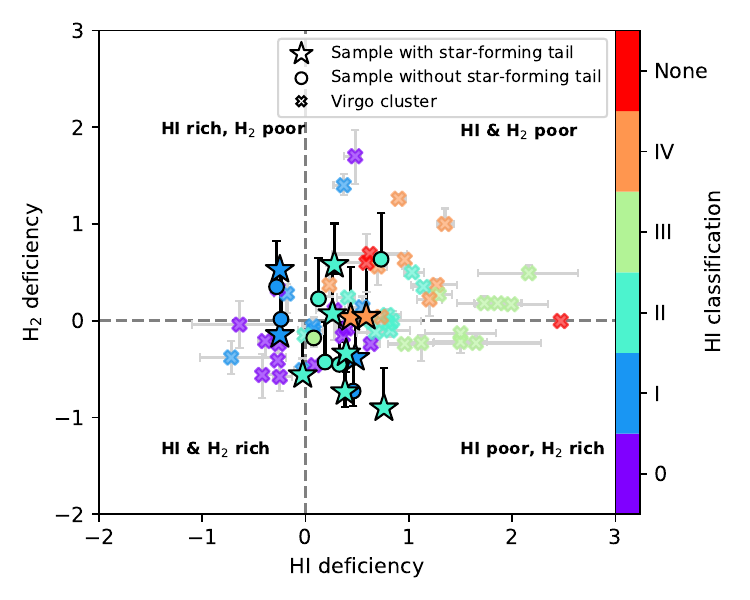}
	\caption{
	Correlation between the \HI and H$_2$ deficiencies. 
	Crosses show the galaxies in Virgo cluster from \cite{Zabel+2022}, while the APEX sample is shown by the colored stars and points, depending on whether the galaxy has star formation in the tail or not.
	Each symbol is colored according to the classification proposed by \cite{Yoon+2017} based on the \HI morphology and the \HI and \hdue content (see text). }
	\label{fig:vertico_deficiency}
\end{figure}

Based on our classification, 
galaxies showing no or very mild \HI deficiency and asymmetries (i.e. class I) have normal or slightly high H$_2$ content compared to field galaxies (see Fig.~\ref{fig:vertico_deficiency}). 
Class IV galaxies, which have low \HI surface densities without clear signs of ongoing or past stripping are usually both \HI and \hdue poor, as shown by the Virgo galaxies (transparent symbols in Fig.~\ref{fig:vertico_deficiency}).
Class IV galaxies in our sample, instead, have a normal \hdue content.
Galaxies displaying clear signs of ongoing ram pressure stripping (class II and III)  have normal or enhanced H$_2$ content.
Clearly, our sample is dominated by class I-II-III, which is expected given the selection criteria.
The comparison between our galaxies and those analyzed in Virgo \citep{Zabel+2022}
is shown in 
Fig.~\ref{fig:vertico_deficiency}.
In general, the APEX galaxies are slightly \HI poor but relatively H$_2$ rich compared to field galaxies \citep[][]{Moretti_2020}, following the same behavior as the class II galaxies in the Virgo sample. 
This result is consistent with the results by \citealt{Zabel+2022}, who concluded that \HI deficiency does not necessarily mean H$_2$ deficiency and that the molecular gas content is affected by ram pressure on different timescales than the atomic gas. 
This suggests that the stripping is less severe on H$_2$ than on \HI, which is likely because the former is denser and more gravitationally bound to the galaxy than the latter \citep{Lee+2017,Boselli+2022,Bacchini+2023}.

\begin{figure}
	\centering
	\includegraphics[width=0.45\textwidth]{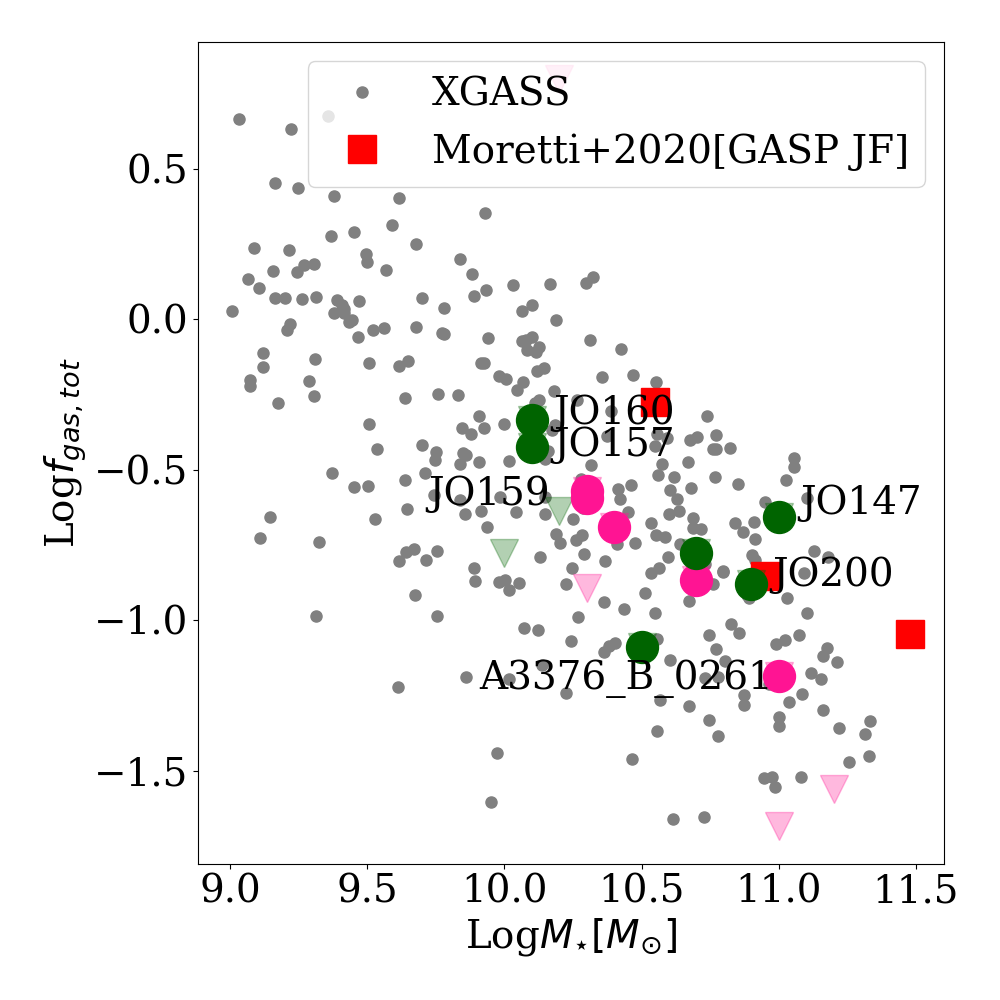}
	\caption{Total cold gas fraction (including both the atomic and the molecular phase) against the galaxy stellar mass for the APEX galaxy sample here described (colors and symbols as in Fig.~\ref{fig:fh2} and the XGASS sample (grey dots) from \citealt{Wang+2020,Catinella+2018}.
	 }
	\label{fig:fgastot}
\end{figure}

However, if we now compute the global gas fraction summing up the contributions of \HI and \hdue, both corrected for the helium content, and compare it with the XGASS sample of field galaxies, we find again (as in \citealt{Moretti_2020}) that the global cold gas content of ram pressure stripped galaxies is pretty normal, as shown in Fig.~\ref{fig:fgastot}.
We stress that our cold gas mass evaluation refers to the global gas mass of each galaxy both for the atomic and the molecular phase, and is therefore minimizing the effects due to the different spatial distribution of the two phases, which can in principle affect the results \citep{Cortese+2016}.
This finding strongly suggests that, for galaxies selected as in this paper based on the presence of \HI asymmetry, and thus going through a specific phase of their transformation in a cluster, an important effect of RPS is the formation of new molecular gas from the \HI that is not stripped.

Combining Figs.~\ref{fig:fh2} and ~\ref{fig:fgastot}, we can see that this conversion of \HI into \hdue within the stellar body (and therefore the simultaneous \HI deficiency and \hdue enhancement) is strongly enhanced when the interaction with the cluster environment leads to having star formation in the tail of stripped gas.

\section{Discussion}\label{sec:conclusions}

If the interpretation proposed in the previous section is correct, then our results should be broadly consistent with the position of the galaxies in the projected phase-space diagram, which is the best observational representation of the galaxy orbital
histories within a cluster. Despite the fact that this diagram is severely affected by projection effects, still it has been used in the past to infer  galaxy properties as a function of their infall history within their host cluster \citep{Rhee+2017,GASPIX,GASPXXI,Franchetto+2021_manga}.
We therefore exploit the phase space diagram for the three clusters containing most of the galaxies here described, i. e. A3558, A85 and A3376 with the purpose of testing this hypothesis. 

\begin{figure*}
\centering
\includegraphics[width=1\textwidth]{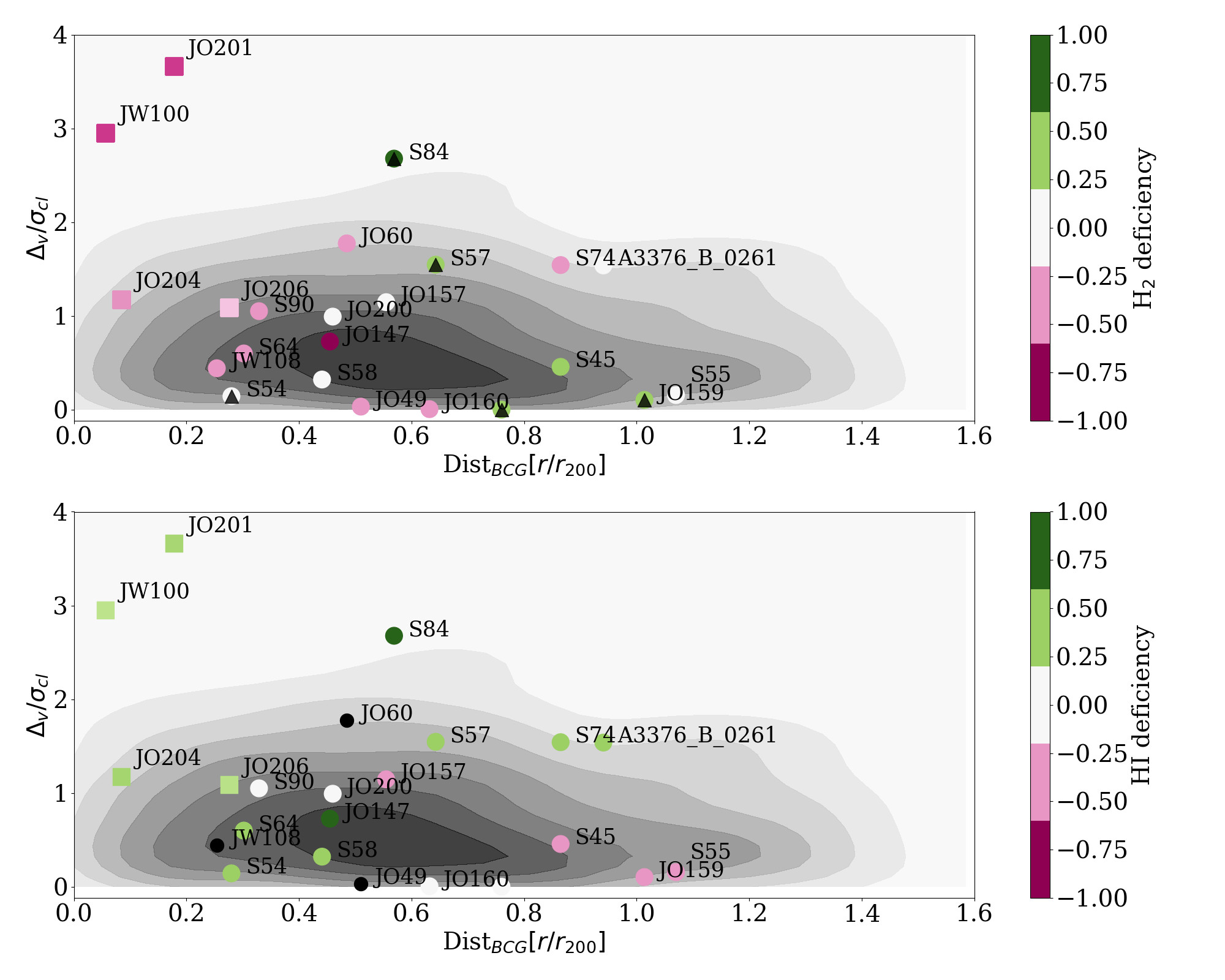}
\caption{
	Phase-space diagram (i. e. galaxy projected velocities normalized to the cluster velocity dispersion ($\Delta_V/\sigma_{cl}$) against the galaxy distances from the central Brightest Cluster Galaxy (BCG) normalized to each custer r$_{200}$) of the 3 cluster (A85, A3558, A3376) members from OmegaWINGS \citep[greyscale contours, from][]{Moretti+2017}, with superimposed the galaxies discussed in this paper (circles) and the 4 GASP galaxies in \cite{Moretti_2020} (squares).
 Galaxies are color-coded according to their \hdue-deficiency (top panel), and to their \HI deficiency (bottom panel), with green symbols being gas deficient and purple being gas rich.
Black triangles in the top panel show low-S/N detections, while black dots in the bottom panel indicate a no detection in MeerKAT (JW108) and missing data (JO60, JO49).}
	\label{fig:ppd}
\end{figure*}

In building this diagram we used all the galaxies confirmed to be cluster members on the basis of the redshift measurements from \citet{Moretti+2017}. The position of each galaxy in the y-axis of the phase-space is given by the projected velocity of each galaxy with respect to the cluster velocity. The position on the x-axis is given by the relative distance of each galaxy with respect to the cluster BCG. In order to build a single phase-space for the three clusters, we normalized the y-axis to the cluster velocity dispersion ($\Delta_V/\sigma_{cl}$) and the x-axis to the cluster r$_{200}$, where both values come from \citet{Moretti+2017}. We plot in Fig.~\ref{fig:ppd} the galaxies number density as greyscale contours.

In Fig.~\ref{fig:ppd} galaxy colors reflect the \hdue deficiency in the top panel, and the \HI deficiency in the bottom panel: in both cases green symbols represent gas poor galaxies, and pink symbols gas rich galaxies.
Dots are the galaxies discussed in the present paper, while squares refer to the four jellyfish galaxies in \citet{Moretti_2020}. When present, the black triangles within the dots indicate low S/N measurements.

As already shown in Fig.~\ref{fig:fh2}, galaxies with an \HI asymmetry in MeerKAT for which a clear APEX detection is present are also \hdue enriched (pink-ish colors in the top panel) and, at the same time, \HI-poor (green colors in the bottom panel). Their position in the phase-space diagram is consistent with the scenario in which galaxies closer to the cluster center appears to show a lower \HI content. The coupling of this effect with the relative \hdue enrichment strongly support the hypothesis that together with some stripping the \HI gas gets also, at least partially, efficiently converted into \hdue, making these objects also \hdue rich.

We notice that this result is compatible with the recent findings 
for Coma, where the atomic phase of cluster galaxies is characterized by a star formation efficiency that is higher than the one shown by their field counterpart
\citep{Molnar+2022}.
In Fornax and Virgo the situation is less clear, as most galaxies do follow normal relations for what concerns the cold gas content and the SFR, with a small percentage of galaxies characterized by a lower \HI content and a higher SFE(\HI) \citep{Loni+2021,Kleiner+2021,Brown+2021,morokuma-matsui+2022}.

Our findings are therefore implying that the galaxies showing this enhanced efficiency SFE(\HI) are exactly those showing a clear asymmetry in \HI and a star forming tail.

\section{Summary}
By analyzing single dish APEX data for a large sample of galaxies with \HI asymmetries, we find that those showing star formation in the tail, either traced by the optical broad band UV, B or \Ha emission, possess within the stellar body an \hdue content which exceeds the value expected for normally star forming field galaxies with the same mass.
On the other hand, those not showing star formation in the tail exhibit a normal/low content of \hdue. 
This result can be explained by an evolutionary sequence in the RPS: first galaxies start losing a negligible amount of their cold atomic gas, and then, when star formation manages to take place in the tail, part of the \HI reservoir in the disk gets efficiently converted in \hdue.
 
In fact, since their global (\HI $+$ \hdue) gas content is not significantly different from the predicted one, then this strongly suggests an anti-correlation between these two gas phases, during the specific phase of galaxy transformation probed by our selection: i.e., that when galaxies possess both an \HI asymmetry, and host SF in the tail of stripped gas.

This finding is evidenced also when the cold gas (atomic and molecular) properties are connected to the infall history of galaxies in the cluster potential, as shown by the evolution of the \HI and \hdue deficiencies in the phase-space diagram.

Although our results are derived from single-dish observations, which do not possess the spatial resolution needed to disentangle the impact of the ram-pressure on the galaxy disks (as done in Virgo by \citealt{Watts+2023} and in GASP galaxies with ALMA data in \citealt{GASPXXII,Moretti_2020}) they still provide a first consistent view on the evolution of the cold gas reservoir in a statistically significant sample of galaxies subject to this environmental effect.

\acknowledgements{
We wish to thank the anonymous Referee for the very constructive report, that helped us improving the paper. 
This project has received funding from the European Research Council (ERC) under the European Union's Horizon 2020 research and innovation programme (grant agreement No. 833824, GASP project and grant agreement 679627, FORNAX project).
B. V., M. G. and R.P. acknowledge the Italian PRIN-Miur 2017 n.20173ML3WW 001 (PI Cimatti). Based on observations collected by the European Organisation for Astronomical Research in the Southern Hemisphere under ESO program 0108.A-0511 (APEX) and 196.B-0578 (VLT/MUSE).  
The MeerKAT telescope is operated by the South
African Radio Astronomy Observatory, which is a facility of the National Research
Foundation, an agency of the Department of Science and Innovation. We
acknowledge the use of the Ilifu cloud computing facility - www.ilifu.ac.za, a
partnership between the University of Cape Town, the University of the Western
Cape, the University of Stellenbosch, Sol Plaatje University, the Cape Peninsula
University of Technology and the South African Radio Astronomy Observatory.
Part of the data published here have been reduced using the CARACal pipeline, partially supported by ERC Starting grant number 679627 “FORNAX”, MAECI Grant Number ZA18GR02, DST-NRF Grant Number 113121 as part of the ISARP Joint Research Scheme, and BMBF project 05A17PC2 for D-MeerKAT. Information about CARACal can be obtained online under the URL: https://caracal.readthedocs.io.
}
\facilities{APEX, VLT(MUSE), MeerKAT, Astrosat} 
\software{GILDAS\footnote{https://www.iram.fr/IRAMFR/GILDAS}\citep{gildas}, Python, Astropy\citep{astropy,astropy2,astropy3}}

\bibliography{references}{}
\end{document}